\documentclass{aa}  

\usepackage{graphicx}
\usepackage{natbib}
\usepackage{subcaption}
\bibpunct{(}{)}{;}{a}{}{,} 
\usepackage{txfonts}
\usepackage[
  breaklinks = true,
  colorlinks = true,
  urlcolor   = blue,
  citecolor  = blue,
  linkcolor  = blue,
]{hyperref}
\usepackage{xcolor}
\usepackage[normalem]{ulem}   
\usepackage{placeins}

\def\Halpha{\mbox{H\hspace{0.1ex}$\alpha$}}
\def\Hbeta{\mbox{H\hspace{0.1ex}$\beta$}}
\def\Heps{\mbox{H\hspace{0.1ex}$\varepsilon$}}
\def\kms{\hbox{km$\;$s$^{-1}$}}

\newcommand{\Ha}{\textnormal{H}\alpha}

\defcitealias{2024A&A...686A.218N}{DNS24}

\begin{document}

    \title{Evidence for plasmoid formation related to a hot UV burst in the solar atmosphere}
    \titlerunning{Plasmoids related to a hot UV burst}

    \authorrunning{M.O. Sand et al.}
    
    \author{Mats Ola Sand
          \inst{1, 2},
          Luc Rouppe van der Voort
          \inst{1, 2},
          Souvik Bose
          \inst{1,2,3,4},
          Daniel Nóbrega-Siverio
          \inst{5,6,1,2,7}, \\
          Jonas Thoen Faber
          \inst{1, 2},
          Reetika Joshi
          \inst{1, 2},
          Ignasi Josep Soler Poquet
          \inst{1, 2}
          }

   \institute{
            Institute of Theoretical Astrophysics, University of Oslo, P.O. Box 1029 Blindern, N-0315 Oslo, Norway
        \and
            Rosseland Centre for Solar Physics, University of Oslo, P.O. Box 1029 Blindern, N-0315 Oslo, Norway
        \and
            Lockheed Martin Solar \& Astrophysics Laboratory, Palo Alto, CA 94304, USA
        \and
            SETI Institute, 339 Bernardo Ave, Mountain View, CA 94043, USA
        \and
            Instituto de Astrofísica de Canarias, 
            38205 La Laguna, Tenerife, Spain
        \and
            Universidad de La Laguna, Dept. Astrofísica, 
            38206 La Laguna, Tenerife, Spain
        \and
        Universit\'e Paris-Saclay, Universit\'e Paris Cit\'e, CEA, CNRS, AIM, 91191, Gif-sur-Yvette, France
            }


 
    \abstract
    {
    Magnetic reconnection in the solar atmosphere drives energetic phenomena such as Ellerman bombs, UV bursts, and surges.
    }
    {
    Our aim is to employ high-resolution observations to advance understanding of such phenomena and their connection between each other.
    }
    {
    We used coordinated observations from the Swedish 1-m Solar Telescope (SST), the Interface Region Imaging Spectrograph (IRIS), the Solar Dynamics Observatory (SDO), and the Hinode/X-ray Telescope (XRT) to analyze an active region with strong and repeated magnetic flux emergence. This region displayed long duration UV burst activity, occurrence of multiple Ellerman bombs, and recurrent ejection of surges. 
    }
    {
    Chromospheric imaging in the \Halpha\ and \Hbeta\ lines show the presence of a dome-like structure that overarched the region where magnetic flux was emerging into the photosphere. The curved chromospheric fibrils were closely associated with bright and similarly curved extensions from a UV burst in the IRIS 1400~\AA\ channel. We observed an episode of formation of a bright sheet along the curved fibrils in \Hbeta\ that subsequently fragmented into plasmoid-like blobs  with peak emission at a Doppler offset of +43~\kms. IRIS \ion{Si}{iv} lines in the same region show complex and non-Gaussian line profiles that display asymmetric extensions to high Doppler offsets. Some profiles have clear components with Doppler offsets in excess of 150~\kms\ on both sides of the nominal line center. The optically thin SDO/AIA channels, including 94~\AA, show intermittent and repeated occurrences of the burst. Emission measure and filter ratio analysis indicate that the burst is multi-thermal and can attain temperatures beyond 1~MK. AIA and Hinode/XRT observations reveal a localized EUV and soft-X-ray counterpart of the burst/dome system, supporting the presence of intrinsically multithermal plasma that reaches coronal, and likely multi-MK, temperatures. 
    }
    {
    Our observations resolved aspects of the interaction between emerging magnetic field and the overlying pre-existing fields. To fully trace the rise of magnetic fields through the atmosphere, and characterise its various forms of associated energy release processes, a continued coordinated effort of acquisition of diagnostics across a wide range of wavelengths is required. 
    }

    \keywords{Sun: photosphere --
              Sun: chromosphere
               }

    \maketitle


\section{Introduction}
\label{introduction}

    Magnetic reconnection has been linked to many energetic phenomena in the solar atmosphere. 
    It is the driver behind solar flares and coronal mass ejections 
    \citep{2022LRSP...19....1P, 
    2015SoPh..290.3425J, 
    2011LRSP....8....1C} 
    on larger scales, as well as Ellerman bombs \citep[EBs;][]{1917ApJ....46..298E, 2002ApJ...575..506G}, UV bursts \citep{2014Sci...346C.315P, 2018SSRv..214..120Y}, surges 
    \citep[e.g.,][]{2008ApJ...683L..83N, 
    2013ApJ...771...20M, 
    2016ApJ...822...18N}, 
    and spicules \citep{2012SSRv..169..181T, 2011ApJ...731L..18M, 2011A&A...535A..95D, 2019Sci...366..890S, 2025A&A...697A.180S, 2026A&A...705A.205S} on smaller scales. 
    Ellerman bombs, UV bursts, and surges often occur in connection with strong magnetic flux emergence episodes 
    \citep[e.g.,][]{2010ApJ...724.1083G, 
    2020A&A...633A..67K}. 

    Recently, \citet[][hereafter \citetalias{2024A&A...686A.218N}]{2024A&A...686A.218N} presented coordinated observations that clearly demonstrate how an episode of photospheric flux emergence at granular scales was followed by an EB, a UV burst, and a surge. 
    The emergence of the magnetic dipole was associated with the appearance of a dark bubble observed in the \ion{Ca}{ii}~K blue wing. 
    This can be explained as a cold dome that results from adiabatic cooling from the expanding magnetised plasma as it rises through the atmosphere
    \citep[see also][]{2014ApJ...781..126O, 
    2015ApJ...810..145D, 
    2016ApJ...825...93O}. 
    More than 10 min after the first appearance of the emerging dipole, an EB occurred where a freshly emerged positive polarity patch appeared to collide with pre-existing negative polarity magnetic fields. 
    The EB was identified as strong wing emission in the \Halpha\ and \ion{Ca}{ii}~K lines. 
    Another 10 min or so later, a UV burst emerged in this region as strong and rapidly varying emission in the IRIS 1400~\AA\ channel. 
    Simultaneously with the UV burst, a surge of cold chromospheric plasma was ejected which reached a maximum extension in the \Halpha\ and \ion{Ca}{ii}~K wing images of 8\arcsec. 

    These energetic events in the lower atmosphere result from magnetic reconnection from the interaction between the emerging magnetic field and pre-existing fields. 
    This leads to significant heating of the local plasma: the observation of UV-burst emission in \ion{Si}{iv} lines and other transition region lines suggest temperatures on the order of 100,000~K. 
    Moreover, the UV burst observed by \citetalias{2024A&A...686A.218N} exhibited a weak EUV counterpart in the SDO/AIA coronal channels suggesting temperatures of around 1~MK.
    These are remarkably high temperatures for the relatively high density region of the upper chromosphere and transition region. 

    The emergence of magnetic flux and its interaction with pre-existing flux leads to formation of current sheets where reconnection can occur, for example along the boundary of the dome structure surrounding the emerging bubble. 
    From theoretical and numerical studies, it is known that current sheets can break-up due to the tearing-mode instability and form plasmoids. 
    The direct detection of plasmoids as magnetic islands in the solar atmosphere is beyond the reach of modern observational capabilities but there have been reports of the observation of highly dynamic small-scale blobs in UV bursts, EBs, and flaring regions that can be interpreted to result from plasmoid formation \citep[see, e.g.,][]{2017ApJ...851L...6R, 
    2021A&A...647A.188D, 
    2023A&A...673A..11R, 
    2024ApJ...966L..29C, 
    2025A&A...696A...3L}. 
    The observation of triangular-shaped \ion{Si}{iv}~1394~\AA\ and 1403~\AA\ line profiles and peculiar profiles with several peaks in UV bursts have also been interpreted as a result of plasmoid formation
    \citep{2015ApJ...813...86I, 
    2020ApJ...901..148G}. 

    \citetalias{2024A&A...686A.218N} observed a rather isolated event of small-scale flux emergence in an ephemeral region.
    Here we expand on this work by analysing coordinated observations of an active region with strong and repeated magnetic flux emergence. 
    This region displayed long duration UV burst activity, occurrence of multiple EBs, and recurrent ejection of surges.
    The IRIS spectrograph slit covered significant part of the UV burst activity which provides a wide variety of complex \ion{Si}{iv} spectral profiles.

        \begin{figure*}
        \centering
          \includegraphics[width=17cm]{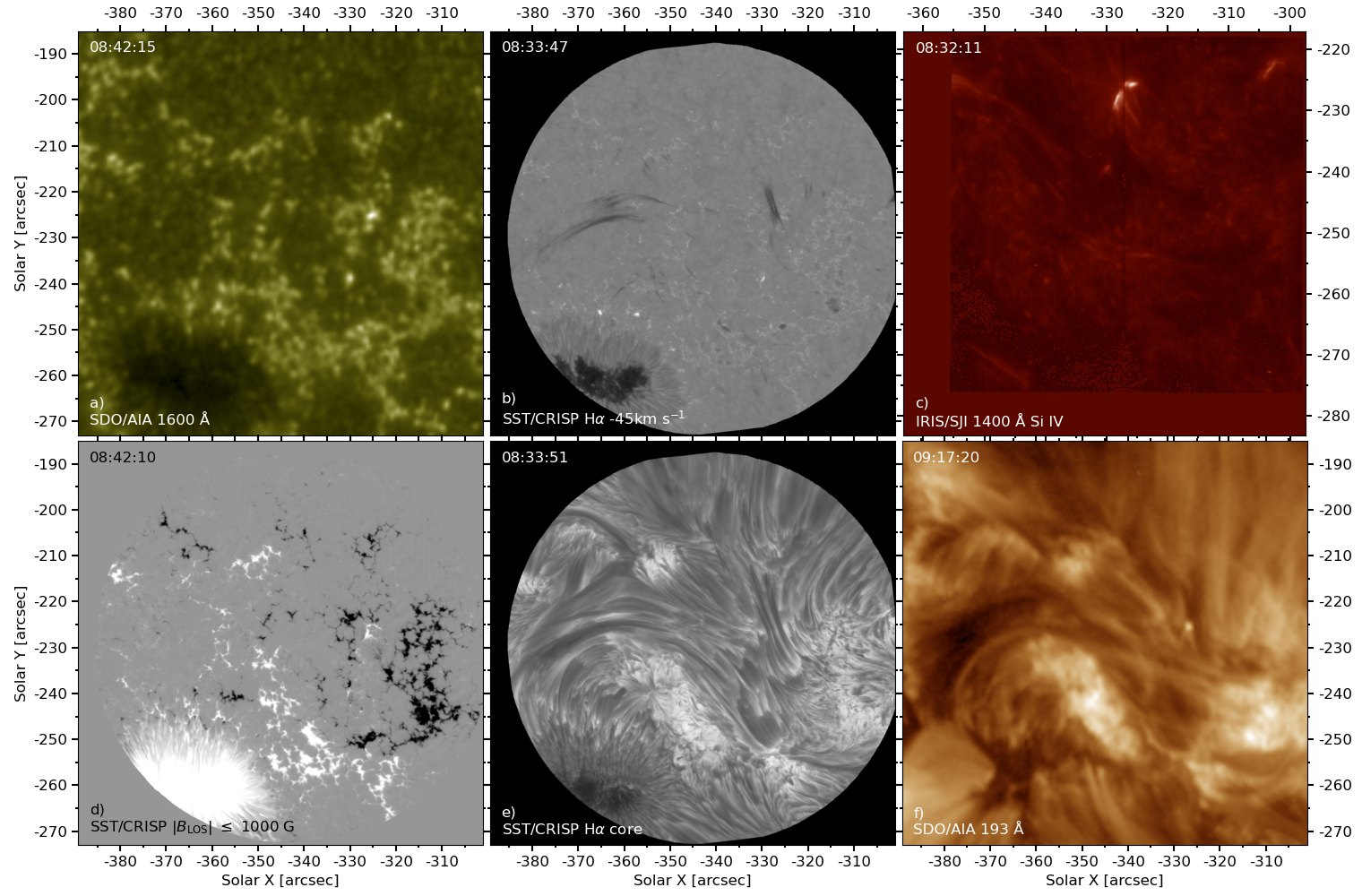}
            \caption{
            Overview of active region AR13380 observed on 27 July 2023. 
            Panel a) shows a cutout of AIA 1600~\AA; panel b) shows an $\Ha$ narrowband image at $-45$~\kms; panel c) shows a SJI~1400 \ion{Si}{IV} image; panel d) shows the magnetic field strength along the line of sight $B_\mathrm{LOS}$ at the photosphere; panel e) shows an $\Ha$ line core image; panel f) shows a cutout of AIA 193~\AA. 
            Panel c) has a different image scale than the other panels, the vertical dashed line marks the position of the IRIS spectrograph slit, and the two vertical dotted lines indicate the area covered by the spectrograph raster. 
            The white or red square indicates the FOV of Figs.~\ref{fig: structure} and \ref{fig: current_sheet}.
            In the associated movie, the AIA 1600~\AA\ and 193~\AA\ panels are affected by calibration procedures during the time interval 08:57 - 09:02. 
            The associated movie is available \href{http://tsih3.uio.no/lapalma/subl/mosand_uvb/mosand_uvb_fig01.mp4}{online}. 
            }
            \label{fig: overview}
        \end{figure*}
        %


\section{Observations}
\label{observations}

    We used coordinated observations from 
    the Swedish 1-m Solar Telescope \citep[SST,][]{2003SPIE.4853..341S}, 
    the Interface Region Imaging Spectrograph \citep[IRIS,][]{2014SoPh..289.2733D}, 
    and the Solar Dynamics Observatory \citep[SDO,][]{Pesnell2012} 
    obtained on 27 July 2023 of active region AR13380 at heliocentric coordinates $(x,y) = (-348\arcsec,-237\arcsec)$ and observing angle $\mu=0.9$.
    This dataset was acquired as part of a multi-season effort to acquire coordinated observations with SST and IRIS \citep[see][]{2020A&A...641A.146R}. 

    IRIS observed between 07:29 and 12:00 UT and was running a medium dense 16-step raster program (observing identifier OBSID3620506835) with the spectrograph slit covering a 5\arcsec$\times$60\arcsec\ area. 
    The exposure time was 4~s and the raster cadence 83~s. 
    Detailed spectral profiles were acquired in several spectral windows, including the \ion{Si}{iv} lines at 1394~\AA\ and 1403~\AA\ (spectral sampling 52~m\AA) and the \ion{Mg}{ii} h and k lines (spectral sampling 25~m\AA). 
    The slit-jaw imager (SJI) was recording images in the 1400~\AA\ passband, dominated by the transition region \ion{Si}{iv} lines, the 2796~\AA\ passband, centered on the chromospheric \ion{Mg}{ii}~k line core, and the 2832~\AA\ passband, in the \ion{Mg}{ii}~h wing and showing the photosphere. 
    The cadence of the SJI images was 20~s. The spatial sampling of both the SJI images and the spectra was 0\farcs166. 

    SST observed for 65 min starting from 08:30~UT. 
    We used the CRISP instrument 
    \citep{2008ApJ...689L..69S} 
    to acquire spectra in the \Halpha, \ion{Fe}{i}~6173~\AA, and the \ion{Ca}{ii}~8542~\AA\ lines at a cadence of 37~s. 
    The \Halpha\ line was sampled at 33 positions between $\pm$2~\AA\ with steps of 0.1~\AA\ between $\pm$1.5~\AA. 
    The \ion{Fe}{i}~6173~\AA\ and \ion{Ca}{ii}~8542~\AA\ lines were observed in spectropolarimetric mode, \ion{Fe}{i}~6173~\AA\ at 13 line positions between $\pm$0.32~\AA\ with 0.04~\AA\ steps in the central part of the line and a continuum position at $+0.68$~\AA. 
    The \ion{Ca}{ii}~8542~\AA\ line was sampled at 15 line positions between $\pm$1.0~\AA\ with 0.07~\AA\ steps in the central part of the line. 
    We acquired a total of 106 time steps with spectral (polarimetric) scans in all three lines. 
    The CRISP FOV was circular with a diameter of about 87\arcsec\ and the pixel scale was 0\farcs044 pixel$^{-1}$.
    We used the CHROMIS instrument to first acquire spectra in the \Hbeta\ line, from 08:30 to 09:05~UT, and then in the \ion{Ca}{ii}~H~3968~\AA\ line, from 09:06 to 09:36~UT.
    The \Hbeta\ line was sampled at 29 line positions between $\pm$2.7~\AA\ with 0.1~\AA\ steps in the central part of the line and at a cadence of 11~s. 
    The \ion{Ca}{ii}~H line was sampled at 44 line positions between $-2.21$ and $+2.34$~\AA\ at a cadence of 17~s. 
    The step size was 0.065~\AA\ in the line core region and the region covering the \Heps\ line at 3970~\AA. 
    The CHROMIS FOV was rectangular with sides 72\arcsec$\times$48\arcsec\ and the pixel scale was 0\farcs038 pixel$^{-1}$.
    The CRISP and CHROMIS data were processed following the standard SST data reduction pipeline 
    \citep{2015A&A...573A..40D, 
    2021A&A...653A..68L} 
    which includes Multi-Object Multi-Frame Blind Deconvolution 
    \citep[MOMFBD,][]{2005SoPh..228..191V} 
    image restoration.
    High data quality was achieved with the aid of the SST adaptive optics system 
    \citep{2024A&A...685A..32S}. 
    We performed Milne-Eddington inversions of the \ion{Fe}{i}~6173~\AA\ Stokes data following the methods developed by
    \citet{2019A&A...631A.153D}. 
    We measured a noise level in $B_\mathrm{LOS}$ of about 8~G in a relatively quiet region in the FOV.
    The SST data are available through the Stockholm SST archive\footnote{\url{https://dubshen.astro.su.se/sst_archive/}}.

    SDO began observing around 08:06 UT after emerging from Earth's shadow. 
    The eclipse had some impact on the observations: the first HMI data were not in focus, and the AIA instrument went through a calibration procedure between 08:56 and 09:02~UT. 
    We aligned the SDO data to the SST observations following the procedure developed by Rob Rutten\footnote{\url{https://robrutten.nl/rridl/00-README/sdo-manual.html}}
    \citep[see Sect. 3 of][for a short discussion]{2020arXiv200900376R}. 
    To study the context and long term evolution of the active region, we made extensive use of the JHelioviewer 
    \citep{2017A&A...606A..10M} 
    tool to explore the SDO data.

    Hinode/XRT \citep{2007SoPh..243...63G} observed the active region during the IRIS observing sequence. We used XRT images in the Al-poly and Be-thin filters to assess the soft X-ray response of the burst/dome system. The XRT observations used in this work were obtained after the SST coordination window but while the long-lived UV burst was still active in IRIS/SJI 1400~\AA. The XRT data were analyzed and co-aligned with the SDO/AIA observations on a common helioprojective grid for the aperture-based comparison described in Sect.~\ref{methods} and Appendix~\ref{appendix:xrt_cut_off}. We used the SunPy \citep{sunpy_community2020} and XRTPy \citep{Velasquez2024} packages for the corresponding data analysis.

    For the alignment between IRIS and SST, we cross-correlated the photospheric IRIS SJI 2832~\AA\ with CRISP wideband data. 
    For analysis of the spectral data from the raster, we downscaled the SST spectral profiles to match the IRIS spatial sampling and produce level 3 data cubes \citep[for a detailed description, see][]{2020A&A...641A.146R}. 
    These level 3 data cubes can be effectively explored with the CRISPEX data browser tool in IDL \citep{2012ApJ...750...22V} 
    and are available in the IRIS database through the public web portal at LMSAL\footnote{\url{https://iris.lmsal.com/search/}}.

    For analysis of the dynamical evolution in the imaging data, we upscaled the SJI images to match the SST spatial sampling
    \citep[see][]{2021A&A...654A..51B, 
    2025A&A...693A...8T}. 
    

\section{Methods}
\label{methods}

        To characterize the IRIS \ion{Si}{iv} spectral profiles, we produced maps of different quantities: peak intensity, Doppler offset of the peak with respect to the nominal line core position, and full-width at half-maximum (FWHM) of the profile. 
        We also fitted single Gaussian profiles to the spectral profiles. Of particular interest are the maps of the $\chi^2$ goodness of fit values to locate non-Gaussian profiles that are for example highly asymmetric or have multiple peaks. These profiles have high $\chi^2$ values. 
        As a measure of asymmetry, we calculated the red-blue asymmetry following
        \citet{2009ApJ...701L...1D}. 
        We first fitted a single Gaussian to the line profile, and determined the
        line centroid of the Gaussian fit. The R-B asymmetry was then calculated by summing over the 50--80~\kms\ offset range 
        \citep{2017ApJ...850..153N}. 

        The \ion{Si}{iv}~1394~\AA\ line is the strongest of the resonance doublet with an oscillator strength that is double of the 1403~\AA\ line. 
        For complex and multiple-peaked spectral profiles associated with UV bursts, it is beneficial to compare the two spectral profiles and consider weak spectral blends from neutral and singly ionised species \citep{2014Sci...346C.315P}. 
        For the \ion{Si}{iv}~1394~\AA\ line we consider the line blends of \ion{Ni}{ii}~1393.33~\AA\ (Doppler offset $-92$~\kms) and \ion{Fe}{ii}~1392.816~\AA\ ($-203$~\kms). For the \ion{Si}{iv} 1403~\AA\ line, we consider \ion{S}{i}~1401.515~\AA\ ($-269$~\kms). 

        We further applied the $k$-means clustering technique to get an overview of the different extreme spectral profiles in the area of the UV burst. Details of this analysis are provided in Appendix~\ref{app:si_iv_spectra}.

        We performed a differential emission measure (DEM) analysis using the \citet{2020ApJ...905...17P} approach to check how much emission associated with the burst stems from plasma at $\geq$1~MK. This code solves for the DEM inversion scheme using the $L$$^{2}$ norm (Tikhonov regularization), which constrains the merit function, aiding its convergence and discarding any physically meaningless solutions \citep[refer to Eq.~14 of][]{2020ApJ...905...17P}. We used the default 31 temperature bin setup from $\log T[\mathrm{K}]=5.5$ to $\log T[\mathrm{K}]=7$ in steps of 0.05 in log space for the inversion scheme.
 
        Furthermore, we also use the popular filter-ratio method \citep{2011SoPh..269..169N} with SDO/AIA data to estimate the coronal temperatures associated with the burst. This technique serves as a consistency check on temperatures estimated from the DEM analysis. The filter-ratio method relies on the near-linear relationship between the ratio of the set of chosen filters of optically thin temperature diagnostics, allowing a unique determination of temperature for each pixel. The caveat, however, is the assumption that the coronal plasma is isothermal. Here we use the ratios of the AIA 171 and 131, and 131 and 94 filters to determine the temperature within $\log T[\mathrm{K}] \in [5.5,6]$ and $\log T[\mathrm{K}] \in [6.75,7.1]$, respectively. The derived ratios and a brief discussion of the technique can be found in Fig.~\ref{fig:appendix_filter_ratio} and appendix~\ref{appendix:filter-ratio}.  \cite{2020ApJ...902...31T} used the 131 and 94 filter-ratio to determine the temperatures associated with microflares, and more recently, \cite{2025ApJ...983L...7B} used the 171 and 131 filters to determine the lower (transition region) temperature cutoff associated with spicules/network jets. 

        To further constrain the coronal contribution of the burst/dome system, we compared the AIA 94~\AA\ emission with co-spatial Hinode/XRT Al-poly and Be-thin emission. The AIA and XRT data were co-aligned and resampled to a common helioprojective grid, and the comparison was performed over a common aperture enclosing the compact burst core and the adjacent dome signature. Following the approach described in Appendix~\ref{appendix:xrt_cut_off}, we converted the AIA 94~\AA\ count rate into an isothermal emission measure for a grid of a given set of temperatures using the AIA 94~\AA\ temperature response function, and then folded this emission measure through the XRT temperature response functions to predict the aperture-integrated XRT signal. The lowest temperature at which the predicted XRT Al-poly flux matches the observed flux provides a conservative lower-temperature consistency constraint. Because the burst/dome system is spatially structured and the XRT filters do not yield a simultaneous single-isothermal solution, this comparison is interpreted as evidence for multithermal coronal-temperature plasma rather than as a unique temperature measurement. This approach is similar to the analysis used by \cite{2025ApJ...983L...7B} using AIA/IRIS to determine the million-K temperature associated with spicules.

        %
        \begin{figure*}
        \centering
          \includegraphics[width=17cm]{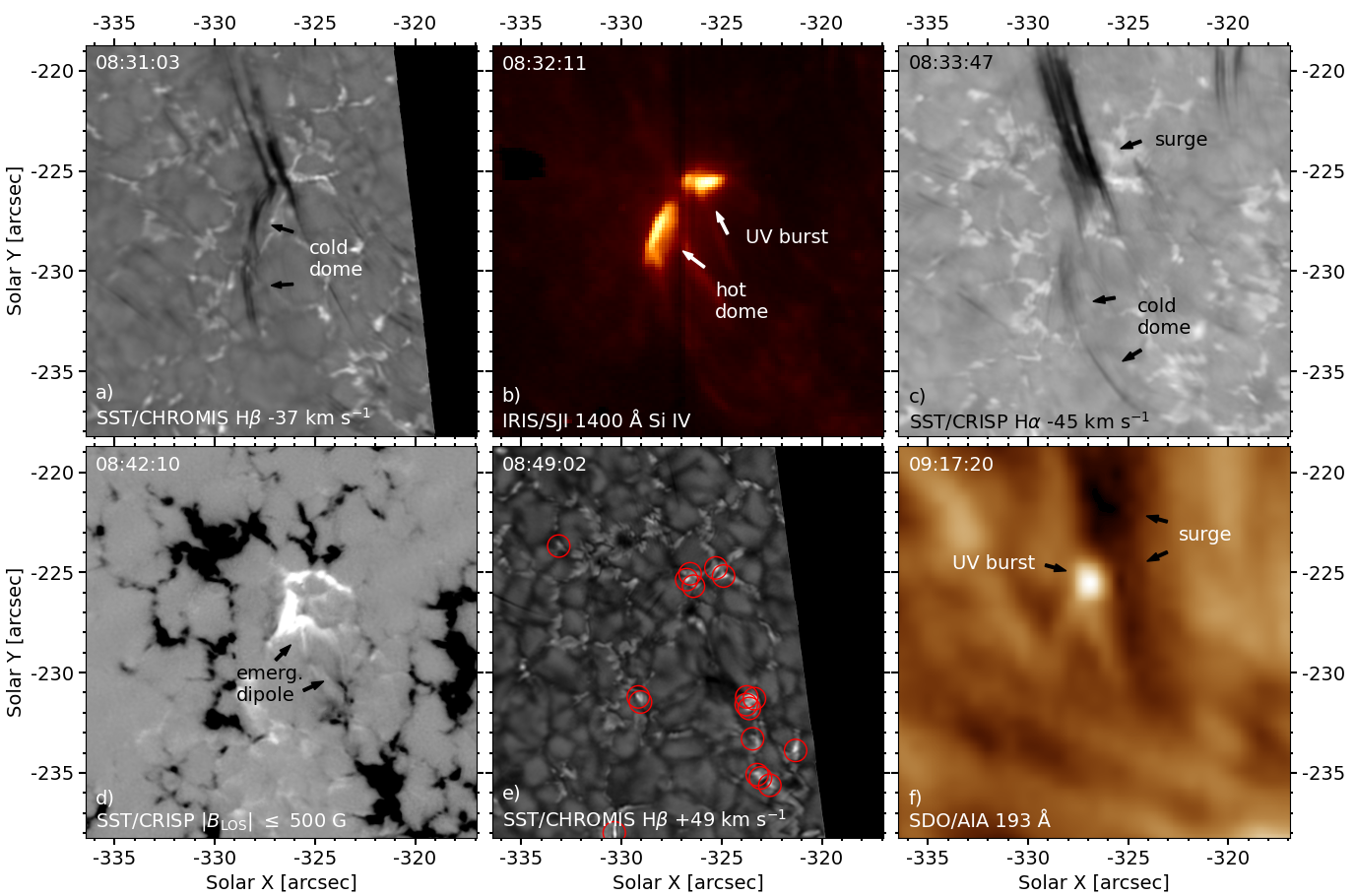}
            \caption{
            Zoom-in on the region with magnetic flux emergence, EBs, a dome structure, a UV burst, and the base of a surge, at different times of the observed time sequence. 
            Panel a) shows the dome structure and parts of the surge in \Hbeta\ blue wing; panel b) shows the UV burst and the hot part of the dome in SJI 1400; panel c) shows the bottom part of a surge in \Halpha\ blue wing; panel d) shows an emerging dipole in a $B_\mathrm{LOS}$ map; panel e) shows EBs in \Hbeta\ red wing, red circles mark regions with enhanced \Hbeta\ wing emission; panel e) shows the UV burst in AIA 193~\AA\, the bright emission from the UV burst is visible next to the dark surge. 
            The associated movie shows the temporal evolution of $B_\mathrm{LOS}$, \Halpha\ blue wing, SJI 1400~\AA, and AIA 193~\AA. 
            The associated movie is available \href{http://tsih3.uio.no/lapalma/subl/mosand_uvb/mosand_uvb_fig02.mp4}{online}.
            }
            \label{fig: structure}
        \end{figure*}

\section{Results}
\label{results}

    \subsection{Overview of energetic events in the observed region}
    
    An overview of the observed active region is presented in Fig.~\ref{fig: overview}. 
    The active region has a large positive polarity trailing spot. The region of interest with EB, UV burst, and surge activity is located in the leading plage region that is dominated by negative polarity magnetic field. 
    The activity occurred in the area where a large positive polarity patch is surrounded by the negative polarity plage (see Fig.~\ref{fig: overview}d). 
    This positive polarity patch is the result of continuous flux emergence and its first appearance started 5.5~h earlier at about 03:15 UT. 
    The UV burst is prominently visible with strong emission in the SJI 1400~\AA\ image and as a strong brightening in the AIA 1600~\AA\ image.
    The \Halpha\ images show the surge: in the \Halpha\ core image in panel e), it has an extension of about 27\arcsec, in the blue wing image in panel b), the extension is about 6\arcsec. 
    In the AIA 193~\AA\ image, bright emission from the UV burst is visible adjacent to the dark surge. 
    In the associated animation, the UV burst is not always visible in AIA 193~\AA\ as it is sometimes obscured by dense surge material. 

    Figure~\ref{fig: structure} presents the different structures in more detail in different diagnostics and at different times in the observed time sequence. 
    The complex structure of the emerging dipole surrounded by different pre-existing negative polarity patches of the leading plage is shown in the $B_\mathrm{LOS}$ map in panel d). 
    The region exhibits strong EB activity: red circles in the \Hbeta\ wing image in panel e) mark locations with strong enhanced wing intensity where the spectral profiles display the characteristic EB mustache-like profile.
    The dark curved fibrils in the \Hbeta\ (panel a) and \Halpha\ (panel c) wing images indicate the presence of a dome structure. 
    The visibility of curved fibrils that outline the dome in these chromospheric diagnostics indicate the presence of dense plasma which constitute colder parts of the dome boundary. 
    The SJI 1400~\AA\ image shows the UV burst and a bright curved extension that closely follows the geometry outlined by the chromospheric fibrils in the \Halpha\ and \Hbeta\ wings. This extension in SJI 1400~\AA\ originates from higher temperature plasma in the dome boundary. 
    The UV burst is also visible in the AIA 193~\AA\ image, as well as a vague extension that seems to outline the dome structure in similar way as in SJI 1400~\AA, but much weaker. 
    The surge that is labeled in the \Halpha\ wing image is also visible in AIA 193~\AA.  

    The animation that is associated with Fig.~\ref{fig: structure} shows the temporal evolution of four diagnostics: $B_\mathrm{LOS}$, AIA 193~\AA, \Halpha\ blue wing, and SJI 1400~\AA. 
    The time sequence of $B_\mathrm{LOS}$ shows a continuous and complex emergence of magnetic flux. It is possible to distinguish at least three distinct flux emergence episodes. 
    In the time sequence of \Halpha\ blue wing, flux emergence episodes are associated with rapid motion of photospheric bright points and EB activity in the form of highly variable and intense brightenings in and around some of the bright points. The $-1$~\AA\ offset (Doppler offset $-45$~\kms) is around the peak of the typical EB spectral profile. Details of the highly variable EB activity is partly hidden in the animation due to saturation from a relatively low value of the upper intensity threshold. 
    The dome structure as outlined by curved chromospheric fibril-like structures is clearly visible in the \Halpha\ wing time sequence. This also appears to be the base from where surge activity is launched. Many separate threads that appear to be part of the surge activity have their lower end at the curved dome structure. 
    In the SJI 1400~\AA\ time sequence, it appears that the bright UV burst is the source of a continuous ejection of curved jet-like extensions. These extensions follow the same curvature as the dome-like boundary in \Halpha\ blue wing. 
    Some of these curved extensions can also be identified in AIA 193~\AA. In this diagnostic, the surge seems to obscure the UV burst for most of the time and the curved extensions appear to be ejected from beneath the surge. The clearest view on the UV burst is between about 09:04 and 09:19 UT
    (also see Fig.~\ref{fig:surge}).

 \begin{figure}
        \centering
        \includegraphics[width=\linewidth]{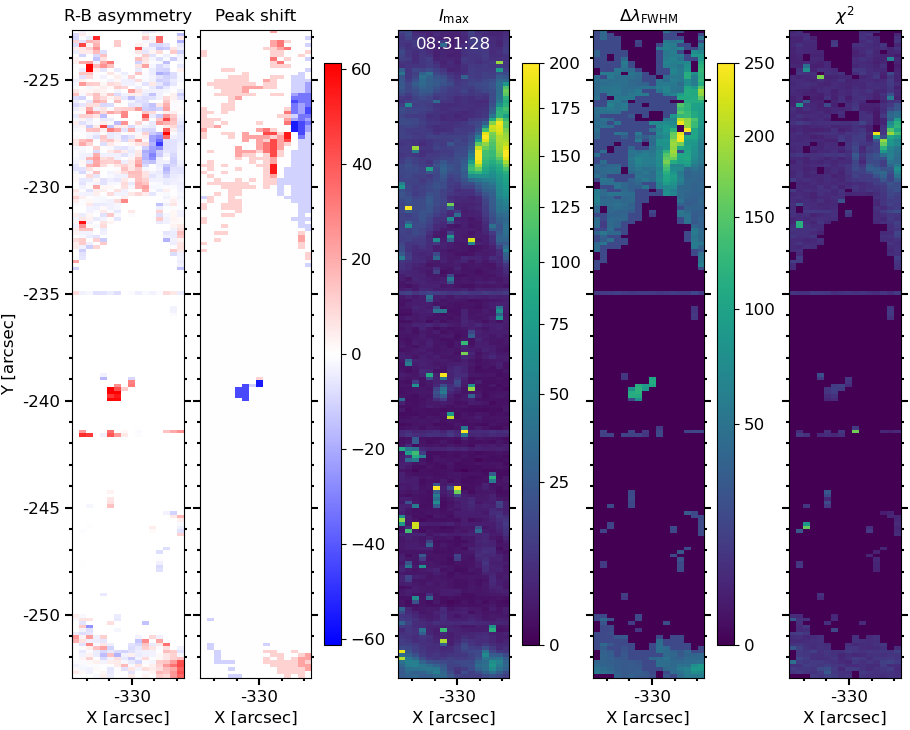}
        \caption{IRIS raster maps of derived quantities of the \ion{Si}{iv}~1394~\AA\ line. From left to right is shown: red-blue asymmetry, Doppler shift of the peak of the profile relative to the peak of average quiet profile (in \kms), peak intensity in counts s$^{-1}$, FWHM in \kms, and $\chi^2$ values of single Gaussian fits. The raster maps consist of 16 slit positions recorded by moving the slit from left to right in about 83~s. 
        This is the first time step in the coordinated time series with SST at raster time 08:31:28~UT. For some pixels, the measurements are affected by noise spikes which can be attributed to the South Atlantic Anomaly.  
        The associated movie is available \href{http://tsih3.uio.no/lapalma/subl/mosand_uvb/mosand_uvb_fig03.mp4}{online}.
        }
        \label{fig:rasters}
    \end{figure}

    \subsection{UV burst}
    \label{res: uv_burst}

        The UV burst was already active at the beginning of the IRIS observations (07:29) and showed clear bursty behavior in SJI 1400~\AA\ before it turned quiet at about 11:05 (see the light curve in Fig.~\ref{fig:appendix_sji1400_lightcurve}). 
        Around this time, the newly emerged positive patch near the UV burst had disappeared, 
        after canceling against the dominant negative polarity magnetic field in the surroundings, and flux emergence in the HMI observations appears to have ceased.
        This gives a lifetime of the UV burst of at least 3~h and 35~min, which means that this UV burst has a relatively long lifetime compared to what is found in the literature \citep[cf.][]{2018SSRv..214..120Y}.
        Most likely, the lifetime of the UV burst was much longer since there was already strong brightening in AIA 1600~\AA\ from the start of the flux emergence at about 03:15~UT, and a clear brightening in AIA coronal channels shortly after. 
        %
        Throughout the lifetime of the UV burst, there are regular periods when the UV burst is not visible in the SDO coronal channels, which appears to be due to obscuration by overlying material from the surge activity. 
        The UV burst was not completely covered by the IRIS raster (see Fig.~\ref{fig: overview}c and associated animation) so a full study of the spatial distribution and temporal evolution of spectral profiles in the UV burst itself is not possible. 
        However, the jet-like extensions emerging from the UV burst are fully covered by the raster. 
        The jet-like extensions reach up to 4\farcs7 in length, while the UV burst itself spans roughly 1\arcsec $\times$2\farcs2 at its largest.
        Figure~\ref{fig:rasters} shows raster maps of different derived quantities from the \ion{Si}{iv}~1394~\AA\ line from a time when there was a strong jet-like extension. 
        The maps of peak intensity ($I_\mathrm{max}$) and line width ($\Delta\lambda_\mathrm{FWHM}$) show the presence of strong and broad \ion{Si}{iv} profiles. The high $\chi^2$ values for these profiles indicate that these profiles are highly non-Gaussian. 
        The Doppler diagnostic maps indicate the complexity of these profiles: the R-B asymmetry and peak shift maps show red and blue patterns that are quite different in the area of the UV burst and the extension. 
        The edge of the UV burst around $(-328\arcsec,-227\arcsec)$ has strongly blue-shifted main peaks of about $-50$~\kms\ while the asymmetry in the wings is skewed towards the red. 
        The extension has strongly red-shifted peaks of more than $+40$~\kms\ with wings that have blue-ward asymmetry. 
        The spectral profiles in the jet-like extension are similar to UV burst profiles: they are strong, broad, non-Gaussian, and have \ion{Fe}{ii} and \ion{Ni}{ii} absorption blends.
        
    \subsection{Signatures of plasmoids along the dome boundary}
    \label{res: dome boundary}

\begin{figure*}
        \centering
          \includegraphics[width=16cm]{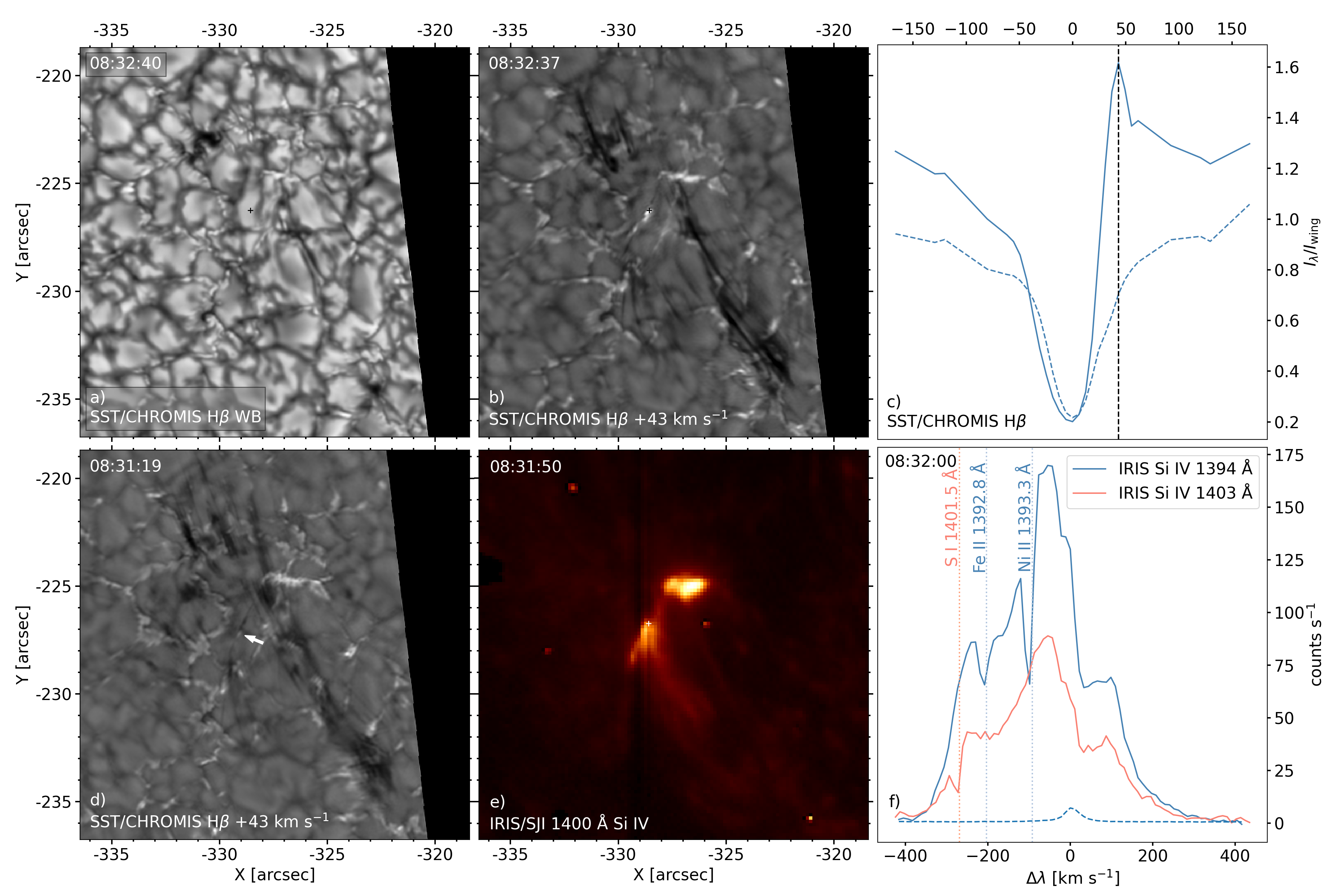}
            \caption{
            Multi-wavelength overview of a possible reconnecting current sheet and plasmoid ejection related to a UV burst. 
            Panel a) shows an \Hbeta\ wideband image, serving as a photospheric continuum reference. 
            Panel b) shows an \Hbeta\ line wing image at $\Delta \lambda = +43$~\kms, at the time with the brightest plasmoid-like blob which is marked with a plus sign. This plus sign (also in panel a) marks the pixel position of the \Hbeta\ spectrum displayed in panel c).
            This panel c) shows the \Hbeta\ spectrum with the strongest peak related to the fragmentation of the current sheet. 
            The blue dashed line represents a reference spectrum, i.e. the average spectrum over space and time. 
            The dashed black line indicates the wavelength position of the \Hbeta\ line wing image.
            Panel d) shows an \Hbeta\ line wing image with a white arrow pointing at the first brightening of the thin, curved structure related to the UV burst. 
            Panel e) shows a SJI 1400 image revealing a narrow, elongated bright structure interpreted as a reconnecting current sheet; the plus sign indicates the pixel position of the \ion{Si}{iv}~1394~\AA\ and 1403~\AA\ spectra displayed in panel f). 
            This panel f) shows the \ion{Si}{iv} spectral profiles, displaying multiple-peaked profiles in both spectra. 
            The pixel index coordinates for spectral profiles is $(t,x,y)=(44, 13, 336)$.
            The vertical lines mark the nominal wavelength positions of weak neutral and singly ionised line blends. 
            The blue dashed spectral profile represents a reference spectrum, i.e. the average spectrum over space and time.
            The peak position of this reference profile serves as Doppler offset reference ($\Delta \lambda = 0$~\kms).
            The associated movie shows the evolution of \Hbeta\ wideband and \Hbeta\ wing.
            The associated movie is available \href{http://tsih3.uio.no/lapalma/subl/mosand_uvb/mosand_uvb_fig04.mp4}{online}. 
            }
            \label{fig: current_sheet}
        \end{figure*}

        The animation associated with Fig.~\ref{fig: current_sheet} shows that the area of the dome boundary is outlined by curved dark fibrils at the beginning of the \Hbeta\ observations from 08:30:56.
        Then, 23~s later at 08:31:19, a slightly bright blob appears in the \Hbeta\ dome boundary, as indicated with a white arrow in Fig.~\ref{fig: current_sheet}d.
        It is located along a very thin curved dark fibril that stretches from about $(x,y)=(-328\arcsec,-226\arcsec)$ to about $(-329\arcsec,-232\arcsec)$. 
        The bright blob brightens in the \Hbeta\ wing and moves about 0\farcs5 upwards along the dome boundary from 08:31:19 to 08:31:52~UT (see the animation associated to Fig.~\ref{fig: current_sheet}). 
        From the next time step at 08:32:04, the blob expands and becomes a bright elongated structure along the curved dome boundary. 
        It reaches maximum extension at 08:32:26 before breaking up into three to four fragments in the next time step (see Fig.~\ref{fig: current_sheet}b), where the clearest blob measures about 0\farcs2$\times$0\farcs4. 
        At this point in time, the \Hbeta\ line profiles in the fragmented brightening possesses strong emission in the red wing, while the blue wing remains unaffected, as shown in the upper right panel of Fig.~\ref{fig: current_sheet}. 
        The \Hbeta\ peak brightening is at about $+43$~\kms\ Doppler offset. 
        We observed a similar brightening in the wing of \Halpha.
        The location of these brightenings is in the middle of a granule in the \Hbeta\ wideband images (Fig.~\ref{fig: current_sheet}a). This implies that the brightenings in \Hbeta\ (and \Halpha) wing images should not be interpreted as photospheric magnetic bright points. It strongly suggests that these bright features are located high above the photosphere, at chromospheric heights. 
        Some bright remnants appear to be visible until 08:33:11~UT, a little more than 30~s after the fragmentation of the elongated brightening.

        During the time of observation of bright blobs and their rapid evolution in \Hbeta, the SJI 1400~\AA\ images show strong brightening of the curved extension (Fig.~\ref{fig: current_sheet}d). 
        The \ion{Si}{iv} profiles in Fig.~\ref{fig: current_sheet}f are from about the same location as the \Hbeta\ profile above but 37~s earlier. 
        The two \ion{Si}{iv} profiles are complex with very strong main peaks that are blue-shifted by about $-25$~\kms. 
        Comparison of the two profiles enables more robust identification of features as possible separate components and avoid ambiguity due to different weak line blends: the red wing is dominated by a component at about $+100$~\kms, there seems to be a component near 0~\kms, and in the far blue wing there is at least one component that is Doppler shifted to more than $-220$~\kms. The intensity ratio of the two lines is close to 2. 

\begin{figure}
        \resizebox{\hsize}{!}{\includegraphics{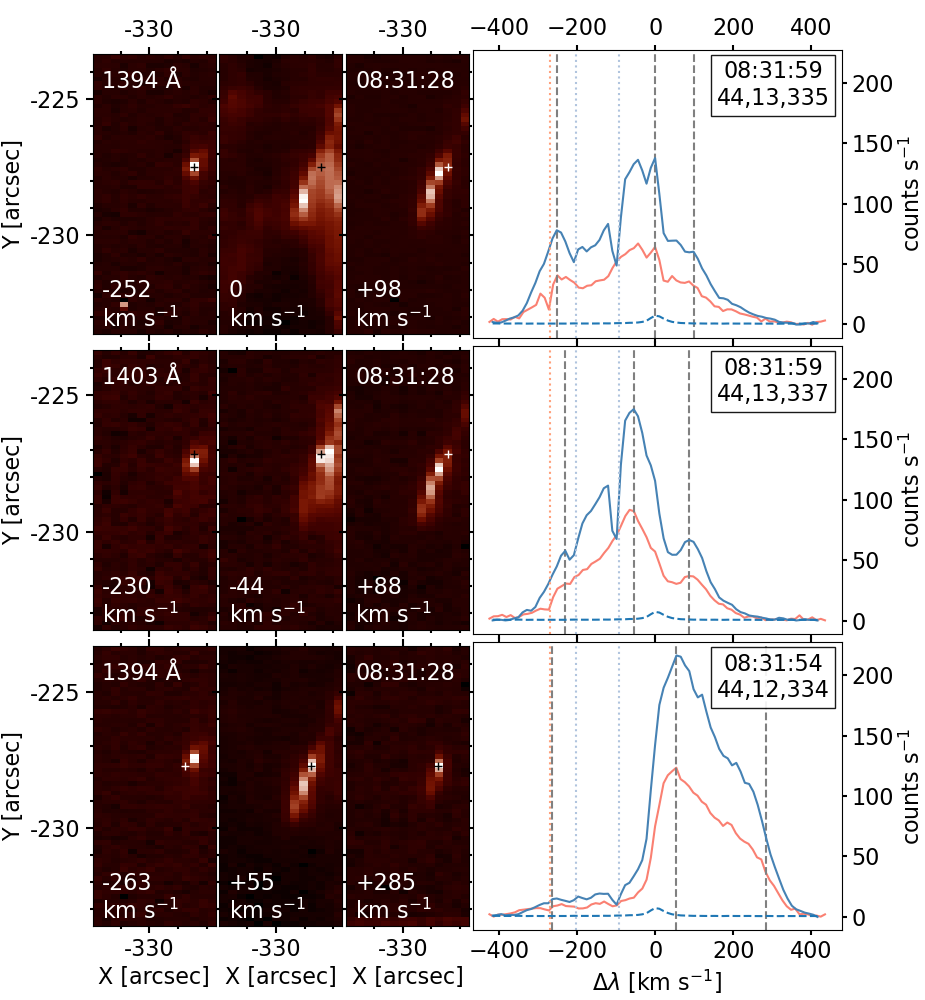}}
        \caption{
            Signatures of plasmoids in IRIS \ion{Si}{iv}~1394~\AA\ and 1403~\AA\ observations. 
            Each row shows simultaneous raster images (three left panels) at different wavelength positions and the corresponding spectral profiles (right panel) extracted from the pixel location marked with the plus sign. 
            Each panel is scaled independently. 
            The blue profile is \ion{Si}{iv}~1394~\AA, the red profile \ion{Si}{iv}~1403~\AA.
            The vertical black dashed lines in the spectral images indicate the wavelength positions of the rasters displayed in each column, following the same order from left to right.
            The vertical, colored, dotted lines mark the nominal wavelengths of the weak line blends; see Fig.~\ref{fig: current_sheet}.
            The average time of the raster images is indicated in the third column, and the timestamp and the pixel index coordinates $(t, x, y)$ of the spectral profiles is shown at the rightmost panel.
        }
        \label{fig: plasmoids}
        \end{figure}

        Raster maps and more \ion{Si}{iv} spectral profiles are shown in Fig.~\ref{fig: plasmoids}. 
        The raster maps are constructed at different Doppler offsets and are all extracted from the same raster scan with an average time stamp of 08:31:28. 
        The raster maps have a blobby appearance and are dominated by single bright pixels or small groups of bright pixels. 
        The raster maps on the left at $-230$~\kms\ and stronger blue-shifts are dominated by two pixels, and the spectral profiles of one of them are shown at the top right. 
        These \ion{Si}{iv} profiles are compatible with a $-252$~\kms\ Doppler shifted component that has an intensity ratio of about 2 for the two lines. 
        This can be interpreted as resulting from a small, unresolved feature moving at extreme velocity, like a plasmoid.  
        This profile also has a component in the red wing.
        %
        The pixel two positions up, for which the spectral profiles are shown in the middle row at right, was recorded at the same time at the same slit position. 
        This profile has an even more pronounced component at $+88$~\kms. 
        In the raster maps around this Doppler shift in the third column, there is no clearly isolated pixel at this location, possibly due to neighboring pixels with higher intensity. 
        For example the nearby bright pixel for which the spectral profiles are shown in the bottom row: it was recorded 5~s earlier and has an extremely bright peak at $+55$~\kms, with more than 200 counts~s$^{-1}$ in \ion{Si}{iv} 1394~\AA, and has strong asymmetry towards the red extending beyond $+285$~\kms.

\begin{figure}[th!]
        \resizebox{\hsize}{!}{\includegraphics{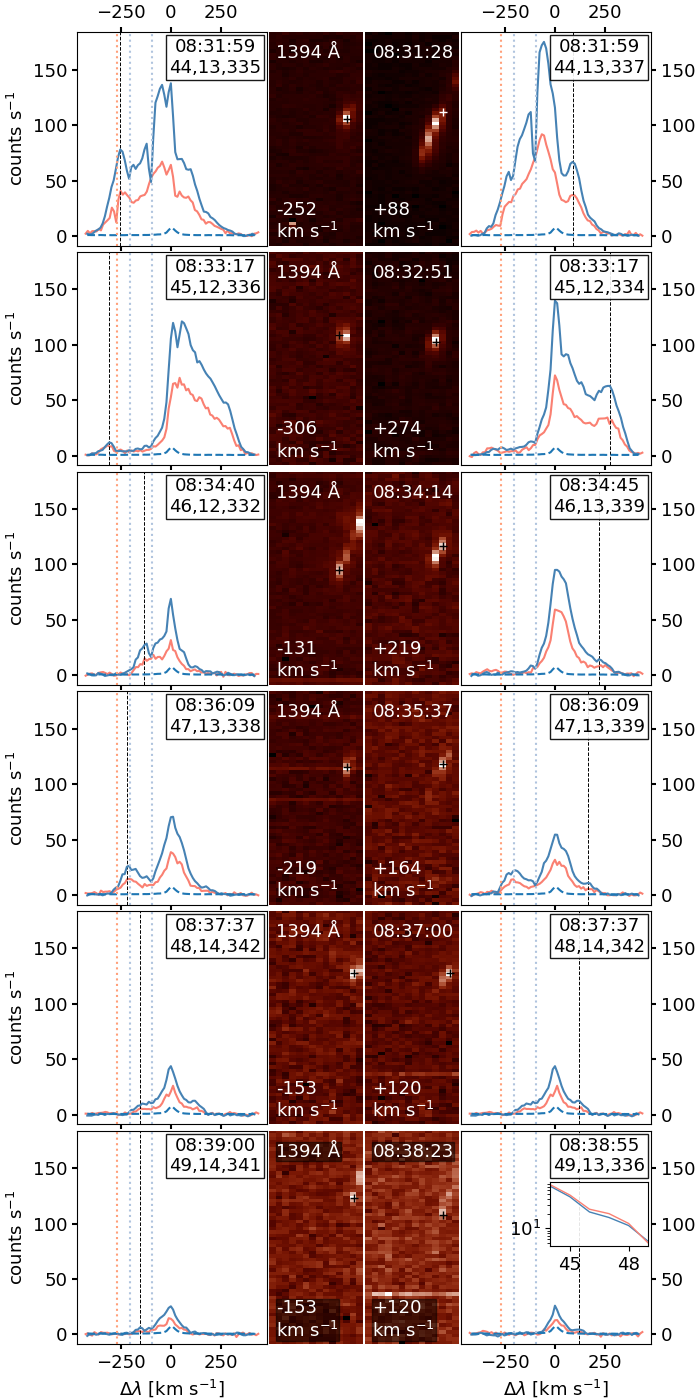}}
        \caption{
            Evolution of plasmoid-like features in \ion{Si}{iv} over six IRIS raster time steps (8 min). 
            The left- and right-most columns show spectral profiles from small bright features in the blue and red wings, respectively (marked with the vertical dashed black lines). 
            Raster maps of these features in the wings are shown in the middle panels with crosses marking the location of the profiles. 
            Each map is scaled individually. 
            The time stamp of the spectral profiles is given at the top of each profile panel, as well as the pixel index coordinates $(t,x,y)$. 
            The time stamp in the raster map panels is the average time of each raster. 
            In the right profile panel in the bottom row, the inset shows the average peak values of the selected features in the blue and red wings over the six time steps in counts~s$^{-1}$ for both \ion{Si}{iv} lines, the $x$-axis is in raster time steps (83~s). 
        }
        \label{fig: plasmoid_evolution}
        \end{figure}

        The raster sequence in Fig.~\ref{fig: plasmoid_evolution} shows more instances of small groups of pixels with spectral profiles that feature additional peaks or strong asymmetries. 
        At 08:32:51~UT, two pixels neighboring the one shown in the top row of Fig.~\ref{fig: plasmoids}, exhibit clear secondary high-velocity components in both \ion{Si}{iv} lines: one pixel shows an isolated peak at $-306$~$\kms$ and another a strong peak at $+274$~$\kms$, indicative of distinct plasmoid ejections.
        At later times, the same region continues to show multi-component profiles. 
        At 08:35:37~UT, the \ion{Si}{iv} spectra display a strong additional component at $-219$~$\kms$, together with an additional weaker component at $+164$~$\kms$ alongside the main peak. 
        Subsequent rasters reveal progressively weaker high-velocity components, with only faint secondary peaks remaining at velocities of approximately $-153$ and $+120$~$\kms$.
        All of the plasmoid-like features, with spatial widths of 0\farcs3–0\farcs7 (2--4 raster pixels), reside within the same area of the dome boundary over 7~min, with secondary peaks appearing in rasters separated by either one (83~s) or two (166~s) scan intervals, suggesting a physical recurrence timescale of at least $\sim$80~s. 
        The speed of the additional peaks decreases from between 250--300~$\kms$ to between 120--150~$\kms$ during this period, while the combined intensity of the blueshifted and redshifted secondary peaks decreases by a factor of $\sim$10, as is shown in the inset plot in the bottom right panel of Fig.~\ref{fig: plasmoid_evolution}.
        Similar evolutions in both intensity and velocity are observed in raster maps associated with other peaks in the SJI~1400~\AA\ light curve (Fig.~\ref{fig:appendix_sji1400_lightcurve}), most clearly for the peak around 08:10~UT, when raster maps reveal a clear, jet-like current sheet.

    \subsection{Coronal response of the UV burst}
    \label{subsec:DEM}

\begin{figure*}
        \centering
        \includegraphics[width=17cm]{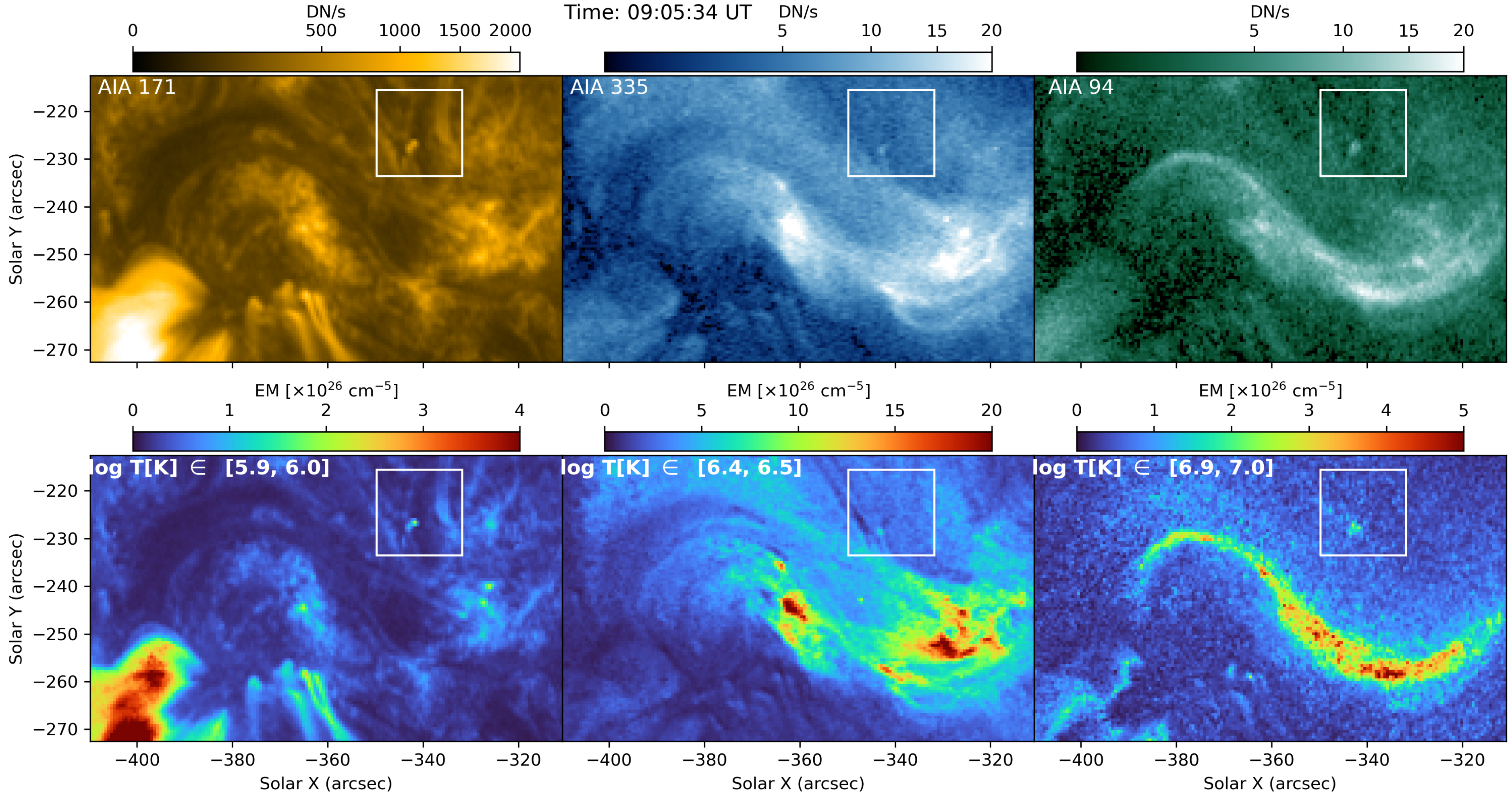}
        \caption{Combined AIA context images (171, 335, and 94 Å; top row) and corresponding emission measure maps (bottom row) for the EUV burst in three temperature bins as indicated. The white box marks the region of interest. 
        The associated movie spans 9 min of solar evolution.
        The associated movie is available \href{http://tsih3.uio.no/lapalma/subl/mosand_uvb/mosand_uvb_fig07.mp4}{online}. 
        }
        \label{fig:AIA_EM_mult_panel}
    \end{figure*}

\begin{figure*}[h!]
        \centering
        \includegraphics[width=17cm]{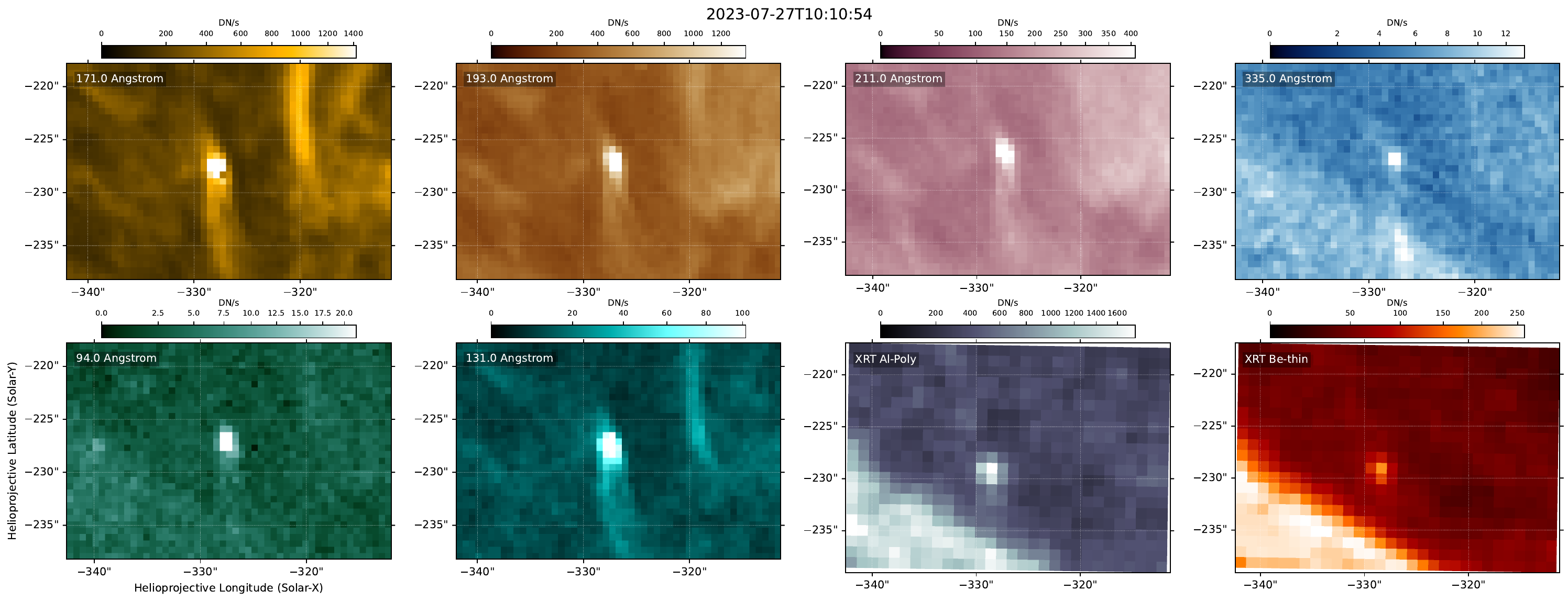}
        \caption{ Zoom-in on the coronal counterpart of the UV burst in coordinated SDO/AIA and Hinode/XRT observations at a time outside the SST coordination window. The top row shows AIA 171, 193, 211, and 335~\AA, and the bottom row shows AIA 94 and 131~\AA\ together with XRT Al-poly and Be-thin. The burst appears as a compact bright core, while a curved extended signature aligned with the dome boundary is traced most clearly in AIA 131~\AA\ and can also be followed in AIA 171, 193, and 211~\AA. Both XRT filters show a clear localized enhancement at the same burst/dome location, supporting the presence of plasma at coronal temperatures.
        The FOV of the zoom-in region is marked in Fig.~\ref{fig:xrt_aia_common}.
        }
        \label{fig:XRT_AIA_burst}
    \end{figure*}

As in \citetalias{2024A&A...686A.218N}, the UV burst also shows an intermittent signature in the optically thin SDO/AIA channels at multiple stages of its lifetime. In the present event, however, the coronal response is stronger, and both the burst and the adjacent dome structure can be identified more clearly in AIA~193~\AA\ (Fig.~\ref{fig: structure}) and in the other coronal channels shown in the top row of Fig.~\ref{fig:AIA_EM_mult_panel}. A zoomed comparison of the burst-centered region in Fig.~\ref{fig:XRT_AIA_burst} further shows that the burst appears as a compact bright core, while a curved extended signature rooted at the burst can be traced along the dome interface. This extension is most clearly visible in AIA~131~\AA, and can also be followed in AIA~171, 193, and 211~\AA, whereas the AIA~94~\AA\ emission is more compact and concentrated closer to the burst core, and the AIA~335~\AA\ response is weaker and more diffuse. Both XRT filters in Fig.~\ref{fig:XRT_AIA_burst} also show a clear localized enhancement at the same burst/dome location, confirming that the event produces a measurable soft X-ray response.

\begin{figure}
        \centering
        \includegraphics[width=\linewidth]{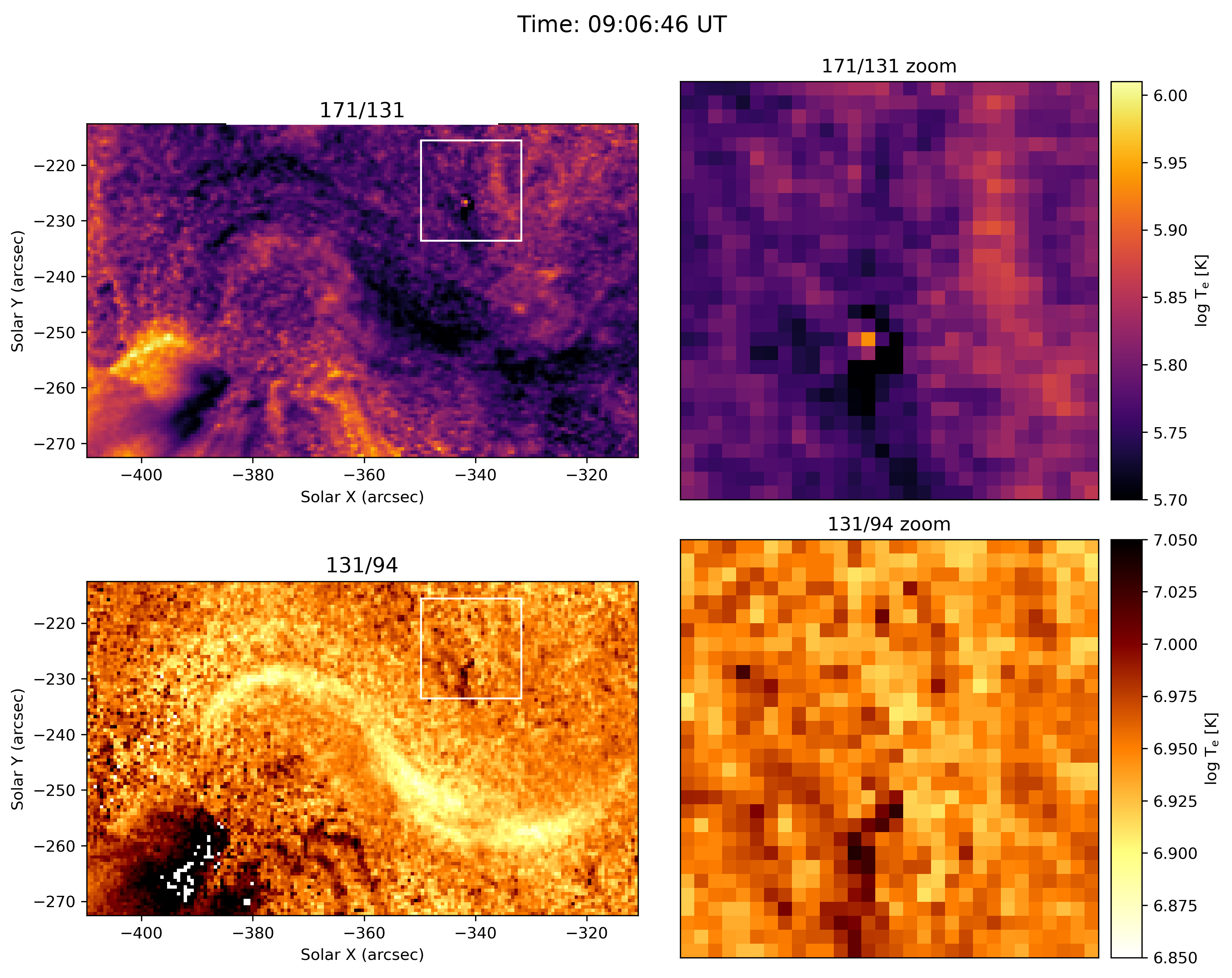}
        \caption{Filter ratio diagnostics to determine the electron temperature ($T_\mathrm{e}$) of the coronal plasma from a pair of SDO/AIA $171/131$ (top row) and $131/94$ (bottom row) channels. The white box marks the region of interest. 
        The associated movie is available \href{http://tsih3.uio.no/lapalma/subl/mosand_uvb/mosand_uvb_fig09.mp4}{online}. 
        }
        \label{fig:filter_ratio_com}
    \end{figure}

The EM maps at $\approx$1~MK, $\approx$2.8~MK, and $\approx$10~MK in the bottom row of Fig.~\ref{fig:AIA_EM_mult_panel} show high emission measure associated with the burst, and signatures that are morphologically somewhat distinct from each other. Such differences imply that the the EUV burst is multi-thermal with temperatures possibly reaching more than a million K. The animation associated with the figure shows the rather intermittent nature of the burst's dynamics and evolution. Figure~\ref{fig:filter_ratio_com} shows the temperature of the EUV burst derived using the filter-ratio method described in Sect.~\ref{methods}. It complements the DEM analysis above as it shows plasma associated with the burst sampling log~$T[K]\approx6$ (top row) and log~$T[K]\approx7$ (bottom row). Interestingly, the figure and its associated animation suggest that the ``hot'' (log~$T[K]\approx7$) response is mainly linked to the extended dome structure immediately below the ``warm'' (log~$T[K]\approx6$) burst.

    Although filter ratios provide a useful and independent consistency check, they do not in general remove the ambiguities inherent in DEM inversions based on the AIA EUV channels, whose temperature responses are broad and multi-thermal. A more technical assessment of the XRT constraint, including the common AIA/XRT field of view and the lower-temperature cutoff inferred from the combined AIA and XRT response, is presented in Appendix~\ref{appendix:xrt_cut_off}. Taken together, the DEM, filter-ratio, and XRT diagnostics support the conclusion that the coronal signature is genuine and that the burst/dome system is intrinsically multithermal. We refer the reader to Sect.~\ref{dis: uv_burst} for further discussion.

\section{Discussion}
\label{discussion}

        \begin{figure*}
        \centering
        \includegraphics[width=\linewidth]{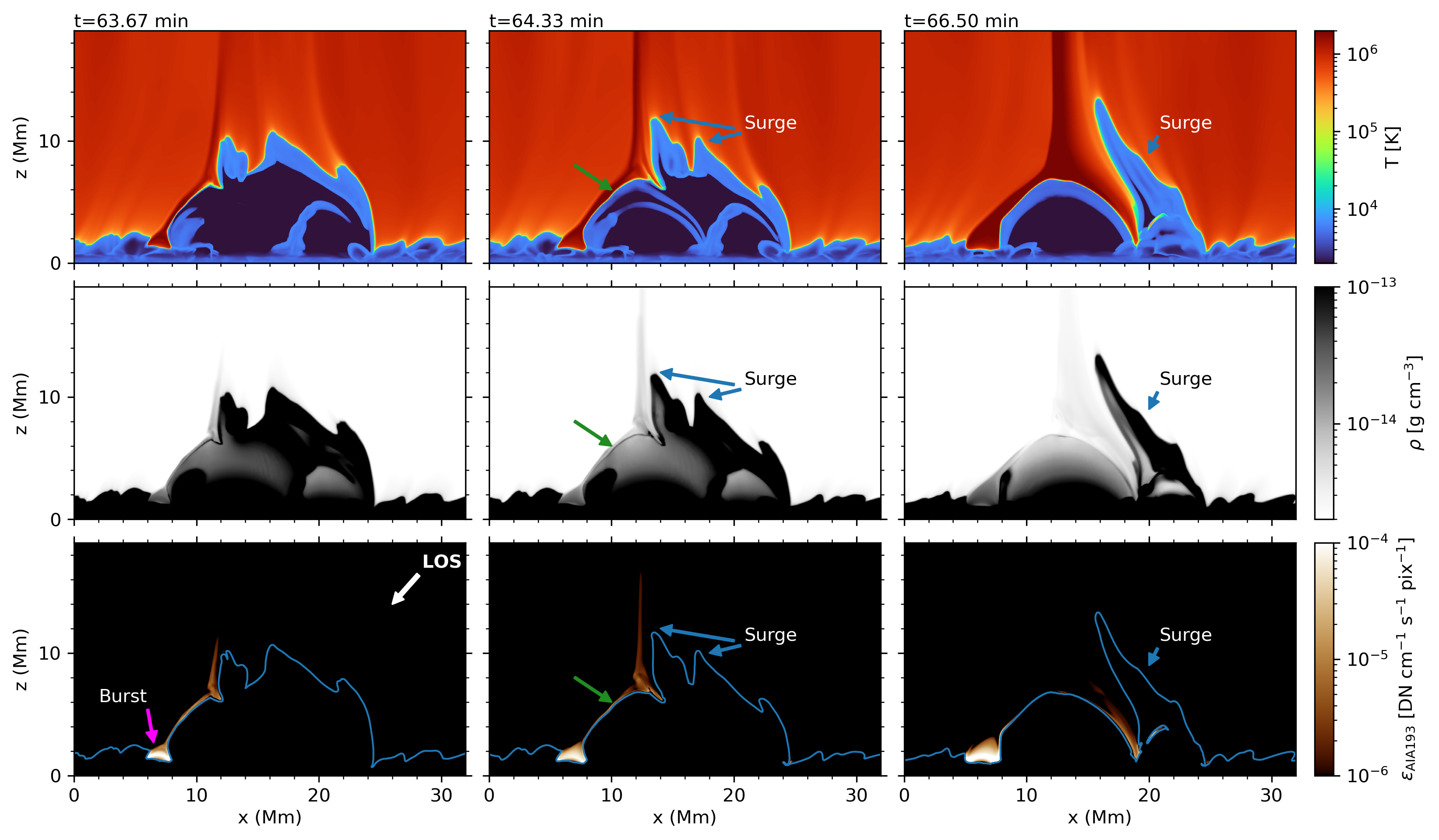}
        \caption{
        Snapshots from a 2.5D Bifrost simulation of magnetic flux emergence in the solar atmosphere.
        The top row shows the temperature, the middle row the mass density, and the bottom row the synthetic AIA 193~\AA\ emissivity.
        The emerging magnetic flux expands into the atmosphere, forming a magnetic dome that interacts with the pre-existing coronal field.
        This interaction produces a cool, dense layer near the upper boundary of the dome, hot plasma immediately above it, and the associated surge and coronal jet.
        The green arrow marks the upper boundary of the dome, where cool, dense plasma and hot emitting plasma coexist in close proximity.
        %
        The blue arrows indicate the cool surge and a pink arrow the location of the burst at the left leg of the dome.
        The white arrow marks a possible line of sight (LOS) for an observer who has the burst blocked by dense surge material for part of the evolution, similar to the observations. 
        }
        \label{fig:simulation}
    \end{figure*}

    We present coordinated observations of an active region that contains an area with repeated episodes of magnetic flux emergence in the leading plage region. 
    HMI observations show that the process of flux emergence started 5.5~h earlier, and AIA imaging show repeated surge activity in this area. 
    There was also long duration (>3.5~h) UV burst activity that was already occurring at the start of the IRIS FUV imaging observations. 
    The UV burst activity ceased at the time when the flux emergence had stopped and the emerged strong positive polarity patch had disappeared as a result of flux cancellation against the dominant surrounding negative polarity magnetic field.
    Chromospheric \Hbeta\ and \Halpha\ imaging show the presence of curved fibrils that outline a persistent dome-like structure that overarched the region where magnetic flux was emerging into the photosphere.
    At times, the curved chromospheric fibrils were associated with bright, aligned, and similarly curved extensions from a UV burst in the IRIS 1400~\AA\ channel. 
    These curved structures emerging from the UV burst can also at times be discerned as weak emissions in coronal AIA channels.  
    We observed an episode of the formation of a bright sheet along the curved fibrils in the \Hbeta\ wing that subsequently fragmented into plasmoid-like blobs. 
    \ion{Si}{iv} spectral lines from the same region show complex and non-Gaussian line profiles that display asymmetric extensions to high Doppler offsets, and sometimes have clear components at Doppler offsets at both sides of the nominal line center wavelength.
    The UV burst itself is at times visible as a compact bright feature in the AIA channels, including AIA 94~\AA. 
    A detailed analysis of the emission in different AIA channels suggests plasma in the UV burst can reach a temperature above 1~MK. 

    \subsection{Flux emergence}
    \label{dis: flux_em}

        The region of interest exhibited emergence of magnetic flux in HMI magnetograms over many hours and this resulted, by the time of the start of the coordinated SST and IRIS observations, in a configuration of a large positive polarity magnetic field patch almost completely surrounded by the dominant negative polarity magnetic fields of the leading plage. 
        While the general pattern of magnetic flux emergence could be clearly observed by HMI, the higher resolution SST magnetograms revealed finer details of the emerging process. At least three distinct dipole emergence episodes happening in close vicinity can be distinguished in the 65~min SST time sequence. 
        The repeated emergence within a confined area points to a preferred site for flux appearance, likely associated with a persistent subsurface magnetic structure.
        Numerical simulations show that buoyant magnetic flux systems rising through the convection zone often emerge in a fragmented and episodic manner due to their interaction with convective flows 
        \citep[e.g.,][]{2004A&A...426.1047A, 
        2007A&A...467..703C, 
        2008ApJ...679..871M, 
        2009A&A...507..949T, 
        2016ApJ...822...18N}. 
        This suggests that the observed events may correspond to successive portions of the same magnetic system reaching the photosphere at different times.

        In the observations of \citetalias{2024A&A...686A.218N}, the more than ten times better resolution of SST allowed the clear identification of the flux emergence episode, while in the HMI magnetograms this episode was indiscernible. 
        In the observations we present here, the recognition of the distinct episodic character of the flux emergence in the CRISP magnetograms, may be connected to the bursty nature with recurrent intensity peaks in the UV burst light curve. 
        The UV burst is caused by magnetic reconnection following the interaction between the newly emerged magnetic field rising from below and the overlying magnetic canopy. 
        A scenario of episodic flux emergence with periods of enhanced rising flux rather than continuous smooth emergence would result in longer term variations in the UV burst light curve on top of short-term variability inherit to the reconnection process. 
        \citetalias{2024A&A...686A.218N} observed a 24~min time difference between the first appearance of the emerging dipole and the onset of the UV burst. 
        This was an isolated flux emergence episode in an ephemeral region so that the connection with the following steps in the chain of events (EB, UV burst, and surge ejection) was clear.  
        It is more challenging to make a direct connection between individual flux emergence episodes and large amplitude variations in the UV burst light curve in the complex active region we study here. 
        It is clear, however, that this UV burst exhibits episodes of enhanced intensity level after an increase of more than a factor 2 over a few minutes, on top of variations on the order of <0.5 (see the light curve in Fig.~\ref{fig:appendix_sji1400_lightcurve}).

    \subsection{Overarching dome structure}
    \label{dis: dome}
        The \Halpha\ and \Hbeta\ spectral imaging reveals a collection of thin, curved chromospheric fibrils associated with the magnetic flux emergence event. 
        Careful inspection of the \Halpha\ and \Hbeta\ spectral imaging at different wavelength positions suggests that these fibrils outline the dome-like structure overlying the emerging region.
        The formation of a dome results from the expansion of the newly emerging magnetic field through the solar atmosphere.
        During flux emergence, the apex of the rising magnetic dome typically retains higher plasma densities than the more evacuated interior
        \citep[e.g.,][]{2001ApJ...549..608M, 
        2005Natur.434..478I, 
        2006PASJ...58..423I, 
        2013ApJ...771...20M}. 
        This layer of enhanced density could be sufficient to produce detectable opacity in \Halpha\ and \Hbeta, thereby delineating the upper part of the dome.
        Moreover, the presence of a surge associated with this flux-emergence episode is consistent with the dome containing relatively cool, dense plasma.

        To support this interpretation, we employ the radiative magnetohydrodynamic (rMHD) simulation of \cite{2017ApJ...850..153N,2018ApJ...858....8N} carried out with the Bifrost code \citep{2011A&A...531A.154G}.
        The setup corresponds to a 2.5D flux-emergence experiment in which the emerging magnetic field gives rise to a hot coronal jet accompanied by a cool surge
        (inspired by the seminal models of \citealp{1995Natur.375...42Y,1996PASJ...48..353Y}, but including a more realistic treatment of the thermodynamics and radiative transfer). 
        Apart from helping to investigate the transition-region emission associated with surges observed with IRIS, this model has also been used to explain plasmoid signatures in chromospheric and transition region lines \citep{2017ApJ...851L...6R}.
        Although the simulation is not intended to reproduce the exact magnetic configuration, viewing angle, or full complexity of the observed event, it provides a useful physical framework for interpreting the relative location of the surge, the upper boundary of the emerged dome inferred from the \Halpha\ and \Hbeta\ observations, and the burst site identified through the AIA 193~\AA\ response \citep[see the appendix of][for details of the emissivity calculations]{2022ApJ...935L..21N}.

        Figure~\ref{fig:simulation} presents three representative snapshots from the evolution of the simulation.
        The left column shows the cold, dense emerged dome, the development of a surge above it, and the location of the burst at the lower-left edge of the dome.
        A line of sight similar to the one illustrated here could explain why in the observations the burst is initially detected in the UV but not in the EUV (see the animation of Fig.~\ref{fig: structure} around 08:35~UT).
        The reason is that the overlying cold, dense plasma can absorb the EUV emission due to the presence of neutral hydrogen and helium for wavelengths $\lambda<912$~\AA\
        \citep{Anzer_Heinzel:2005}.
        Figure~\ref{fig:surge} in Appendix~\ref{appendix:surge} shows different time steps in the observed time series, illustrating how the burst becomes visible in AIA~193~\AA\ at later times. 
        This figure also includes images in \Halpha\ line core where the surge is more extended than in \Halpha\ and \Hbeta\ wing images. The surge appears to cover a large fraction of the dome in \Halpha\ line core, just like in AIA~193~\AA.
        From the line of sight in Fig.~\ref{fig:simulation}, the surge in the simulation also covers the dome. 
        As the simulation evolves, the surge develops crest-like structures, highlighted in the middle column of Fig.~\ref{fig:simulation}, which propagate away from the burst in a manner reminiscent of those seen in the animation of the Fig.~\ref{fig: structure} around 09:07 UT (the progressive ejection of surge material would also explain why the burst becomes partially visible in AIA 193~\AA\ in this movie too).
        At the same time, the upper part of the dome, which contains a thin, curved layer of enhanced density (green arrow), becomes exposed.
        The observer may then simultaneously detect the upper part of the cool dome and a hotter region immediately above it with enhanced AIA 193~\AA\ emission, closely resembling the morphology the observations (see the animation of Fig.~\ref{fig: structure} around 09:20 UT, and Fig.~\ref{fig:surge}).
        At later stages, in the right column of Fig.~\ref{fig:simulation}, the surge is completely detached from the emerged region.
        
        Although the interpretation presented above seems plausible, we cannot completely rule out alternative scenarios, as the nature of small-scale eruptions and fine chromospheric structures is complex and often remains ambiguous.
        For instance, flux emergence regions often appear as systems of very dark, low-lying filaments known as arch filament systems \citep[AFS;][]{Gonzalez-Manrique_etal:2018}, which can give rise to jets through peeling-like mechanisms \citep[][]{Joshi2024}.
        In addition, small-scale filamentary structures or minifilaments may form through flux cancellation and subsequently erupt, driving jet-like activity and localized brightenings
        \citep[e.g.,][]{2015Natur.523..437S, 
        Panesar_etal:2017,Panesar_etal:2022}.
        Numerical models have explored this class of events at large scales, in which a strongly sheared filament channel or flux-rope-like magnetic structure becomes unstable and erupts \citep[e.g.,][]{Wyper_etal:2017,Wyper_etal:2018}. 
        Such eruptive behavior is not limited to cancellation-driven evolution: flux-rope formation and eruption have also been obtained in 3D flux-emergence experiments \citep[e.g.,][]{2013ApJ...771...20M,Archontis_Hood:2013} and supported by observations \citep[][]{Joshi2020}. 
        Another possibility is that chromospheric fibrils themselves become unstable and erupt, producing cool ejecta whose morphology may resemble that of a miniature filament eruption \citep[][]{Nobrega-Siverio_etal:2023}.

    \subsection{Plasmoid formation}
    \label{dis: plasmoids}

    The bright extensions emerging from the UV burst may appear as jet ejections when inspected only in the SJI 1400~\AA\ time sequence.  
    However, analysis of the \ion{Si}{iv} spectral line profiles shows that these extensions do not harbor simple plasma outflows that one might associate with jets. 
    Raster maps of this region show both red- and blue-shifts of the main peak, and the line asymmetry measurements are at times showing Doppler asymmetry in opposite direction to the main peak offset. 
    Most importantly, the spectral profiles in the bright dome area are complex, non-Gaussian, and sometimes exhibit clear multiple components at high Doppler offset. 

    Rather than a jet ejection, it is more plausible that these bright extensions are part of a reconnecting current sheet. 
    At the location of a bright extension in SJI~1400, we observe along the curved fibril in \Hbeta, first the formation of a small bright blob, that moves and expands, then forming an extended bright sheet before breaking up into smaller fragments. The whole sequence lasted just below 2~min, including the fading of the last blob. 
    In the brightest blob fragment, the \Hbeta\ line profile shows a strong emission enhancement with a peak at about $+43$~\kms. 
    This sequence of events follows the theoretical picture of tearing-mode instability in an elongated current sheet that breaks-up into plasmoids. 

    Further evidence for the formation of plasmoids comes from the \ion{Si}{iv} profiles. 
    The profiles we observed here are similar to UV burst profiles that were compared to simulations of plasmoid-mediated magnetic reconnection
    \citep{2015ApJ...813...86I, 
    2017ApJ...851L...6R, 
    2020ApJ...901..148G}. 
    Apart from examples of asymmetric profiles with extensions in the wings beyond Doppler offsets of $\pm150$~\kms, we also find profiles with clear components in, for example, the blue wing at Doppler offsets $-306$ and $-252$~\kms, and in the red wing at $+88$ and $+274$~\kms. 
    The intensity ratio of the two \ion{Si}{iv} lines for these components is close to 
    2, which strongly suggest that they originate from small, optically thin, structures moving at extreme velocity along the line of sight. 
    The raster maps at different Doppler offsets also show small bright features of only a few spatial pixels wide.

    The observed quasi-periodic recurrence of plasmoid production, with separations of 83--166~s, is consistent with theoretical predictions of $\sim$1--2 min timescales for plasmoid-mediated reconnection in UV bursts \citep{2019A&A...628A...8P}. 
    Our 83~s raster cadence provides a lower limit on the true plasmoid formation 
    timescale; faster recurrence would remain undetected.
    The systematic decrease in both intensity (by a factor of $\sim$10) and velocity (from 250--300~$\kms$ to 130--150~$\kms$) over 7~min indicates progressive depletion of available magnetic energy. 
    This behavior is consistent with episodic flux emergence providing finite energy for reconnection with the overlying canopy. 
    Each emergence episode yields magnetic energy that is progressively exhausted through plasmoid production, as reflected in the rise and decline of UV burst intensity in the SJI~1400~\AA\ light curve (Fig.~\ref{fig:appendix_sji1400_lightcurve}). 
    This mechanism appears to sustain reconnection throughout the 3~h and 35~min duration of the UV burst.
    \subsection{(E)UV burst}
    \label{dis: uv_burst}

    The DEM analysis of the burst reveals strong emission at temperatures up to $\log\,T\,[\mathrm{K}]\approx 6.9$. Although the presence of $\gtrsim$1~MK plasma in UV bursts remains debated \citep[e.g.,][]{2014Sci...346C.315P, 2022A&A...657A.132K}, $\geq$1~MK signatures are frequently reported in association with UV bursts in both coronal imaging \citep[e.g.,][]{2018ApJ...856..127G, 2024A&A...686A.218N} and spectroscopic \citep{2019ApJ...871...82G} observations. In the present event, the evidence for a coronal counterpart is particularly compelling: the burst is clearly detected in AIA~94 and 131~\AA, and it is also visible in both XRT filters, Al-poly and Be-thin (albeit at a different time but is a part of the same burst/dome system). Taken together, these diagnostics strongly support the conclusion that at least part of the burst/dome system is heated to coronal temperatures. MHD simulations \citep[e.g.,][]{2017ApJ...839...22H, 2019A&A...626A..33H} suggest that burst-associated current sheets can extend over several chromospheric scale heights and intermittently heat plasma to $\geq$1~MK and can therefore have enough emission to appear in the relatively hotter AIA channels.

    A key practical limitation, however, is that DEM inversions based on AIA data alone can overestimate the amount of high-temperature plasma ($\log\,T\,[\mathrm{K}]\geq 6.4$) by at least a factor of three \citep{2024ApJ...961..181A}. Although electron temperatures inferred from filter ratios are broadly consistent with the DEM results, supporting a multithermal nature of the (E)UV burst, both diagnostics have well-known limitations \citep{2020ApJ...905...17P,2011SoPh..269..169N}, including contamination from cooler ions in the AIA/EUV channel response functions. For this reason, the coordinated AIA/XRT comparison presented in Appendix~\ref{appendix:xrt_cut_off} provides a more direct constraint on the thermal nature of the burst. Unlike the AIA-only DEM and filter-ratio analysis, it compares the burst in a common AIA/XRT field of view and uses the different AIA and XRT temperature responses to test whether the AIA~94~\AA\ signal can be explained mainly by lower-temperature plasma. If too much of the AIA burst emission is assigned to low temperatures, the predicted XRT Al-poly flux exceeds the observed value, thereby placing a meaningful lower limit on the emitting plasma. At the same time, the lack of a simultaneous single-isothermal solution for Be-thin indicates that the burst/dome system is not well described by a single temperature component. Taken together, these results strengthen the interpretation that the event is intrinsically multithermal, with the strongest heating concentrated near the compact burst core and additional coronal-temperature plasma extending along the dome interface.

        \subsection{Concluding remarks}
        \label{sec:concluding}

        The observations we present here and in \citetalias{2024A&A...686A.218N} result from a multi-year coordinated effort targeting flux emergence regions to study the interaction of the rising magnetized plasma with pre-existing magnetic fields and the associated energetic phenomena. 
        Achieving full coverage of the whole region of interest and complete evolution with all instruments is not trivial. For example, the region with the UV burst in \citetalias{2024A&A...686A.218N} was not covered by the IRIS spectrograph slit. The IRIS raster was better positioned for the observations we present here but still only partly covered the UV burst itself. The pointing of telescopes in space typically require planning time lines of more than one day which makes covering targets of about 10\arcsec\ size and that evolve on time scales of hours challenging. Another major limiting factor is atmospheric seeing affecting the ground-based observations. The acquisition of $>1$~h high quality data with SST happens only a few times per observing season. 

        \citetalias{2024A&A...686A.218N} analysed an isolated small-scale flux-emergence episode in an ephemeral region. The emergence was followed by an EB, an $\sim$18~min duration UV burst, and a surge. Here we analyse the evolution of longer duration flux emergence in a large active region. During the time of the SST observations, the process of flux emergence was resolved in distinct episodes of emergence of dipoles in close proximity. It was associated with high EB activity with occurrence of multiple isolated and clusters of EBs. The UV burst activity was of remarkably long duration and there was continuous surge activity. The presence of plasmoid-like blobs in UV bursts have been reported before \citep{2017ApJ...851L...6R}. Here we report evidence of plasmoid formation further away from the UV burst, in the current sheet at the dome boundary which forms the interface between the emerging new flux and the overlying pre-existing flux.   

        In contrast to earlier reports of weak EUV counterparts, the event shows an intermittent but distinct response in optically thin AIA coronal channels including 94~\AA. DEM and filter-ratio diagnostics indicate a multithermal nature with plasma reaching at least $\gtrsim$1~MK, with the hottest signatures preferentially associated with the extended dome boundary rather than the compact burst core. However, the estimation of temperatures from the SDO/AIA channels remain ambiguous due to presence of ``cool'' contaminants in the channels' temperature response. To resolve this issue, future observations with the upcoming Multi-slit Solar Explorer \citep[MUSE,][]{2020ApJ...888....3D} will be needed in coordination with SST and IRIS. MUSE's high-cadence multi-slit sub-arcsecond coronal imaging spectroscopy will be ideally suited to quantify how frequently UV-burst/surge systems produce coronal-temperature plasma and to resolve the associated reconnection outflows and heating on their intrinsic timescales.

\begin{acknowledgements}
The Swedish 1-m Solar Telescope (SST) is operated on the island of La Palma
by the Institute for Solar Physics of Stockholm University in the
Spanish Observatorio del Roque de los Muchachos of the Instituto de
Astrof{\'\i}sica de Canarias.
The SST is co-funded by the Swedish Research Council as a national research infrastructure (registration number 4.3-2021-00169).
IRIS is a NASA small explorer mission developed and operated by LMSAL with mission operations executed at NASA Ames Research center and major contributions to downlink communications funded by ESA and the Norwegian Space Centre.
This research is supported by the Research Council of Norway, 
project numbers 325491, 
358540, 
360813, 
and 361159, 
and through its Centres of Excellence scheme, project number 262622.
S.B. acknowledges support from NASA's IRIS (NNG09FA40C) and MUSE (80GSFC21C0011) contracts to LMSAL.
D.N.S. acknowledges support by the European Research Council through the Synergy Grant number 810218 (``The Whole Sun'', ERC-2018-SyG) and by the Centre National d'Etudes Spatiales (CNES) Solar Orbiter project.
The simulations were performed on resources provided by
Sigma2 - the National Infrastructure for High-Performance Computing and Data Storage in Norway.
This research is also supported by the International Space Science Institute (ISSI) in Bern, through ISSI International Team project \#535 Unraveling surges: a joint perspective from numerical models, observations, and machine learning.
We made much use of NASA's Astrophysics Data System Bibliographic Services. 
\end{acknowledgements}


\begin{appendix}

    \section{Categorising \ion{Si}{iv} spectral profiles}
    \label{app:si_iv_spectra}

        To obtain a comprehensive overview of the different spectral profiles in the UV burst and hot dome, we used SciPy's \citep{2020NatMe..17..261V} version of the $k$-means machine learning technique on the IRIS \ion{Si}{iv} raster data. 
        The $k$-means algorithm separates $m$ data points with $n$ features into $k$ clusters, by minimising the Euclidean distance between each feature point of the data points and the cluster centroids \citep{1056489, e4fc5e3f-57ed-3636-9789-4190a319a2fa, zbMATH03340881}. 
        In our case, $m$ is the number of pixels in the region of the UV burst, while $n$ is the number of wavelength points along the \ion{Si}{iv}~1394~\AA\ spectral line.
        The $k$-means algorithm has been used comprehensively in solar physics for classification of different profiles, e.g., UV bursts \citep{2022A&A...657A.132K}, microflares and flares \citep{2023ApJ...956...85T, 2026A&A...705A.174T}, surges \citep{2021A&A...655A..28N}, spicules \citep{2019A&A...631L...5B, 2021A&A...647A.147B, 2021A&A...654A..51B}, and Stokes profiles \citep{2023A&A...675A.130M}. 
        It has further been used for automated feature detection of for example quiet Sun Ellerman bombs \citep{2020A&A...641L...5J, 2022A&A...664A..72J, 2024A&A...689A.156B, 2025A&A...697A.180S}. 
        Our motivation for classifying the profiles is to facilitate the identification of peculiar triangular-shaped and/or multi-peak profiles that are associated with UV bursts. 
        To make sure the training set was not dominated by the quiet background, and since the emission strength of the profiles related to the UV burst is much stronger than the surroundings, we only chose pixels related to the area of the UV burst, plus an extra 10 pixels above and below the event as a margin of error.
        We further restricted the time window for the analysis to the period that was observed with SST. 

\begin{figure*}[h!]
        \centering
        \includegraphics[width=\linewidth]{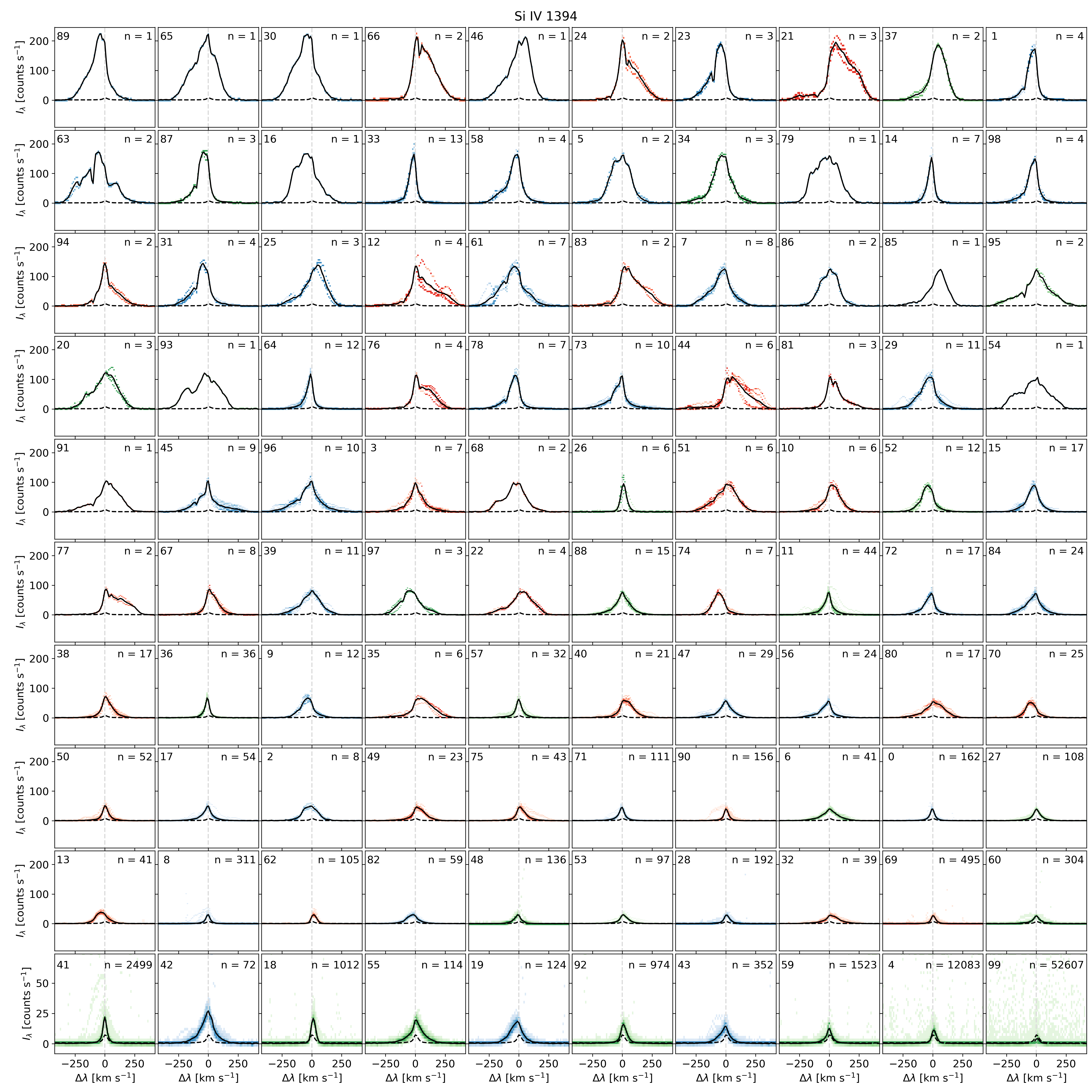}
        \caption{Representative profiles for all 100 clusters in the $k$-means clustering. The cluster panels are organized in order of peak intensity, starting from the top left (cluster numbers from the internal numbering are given in the top left panel corners). The average quiet reference profile is shown as the dashed line. The number of profiles $n$ in each cluster is given in the top right of each panel. The data points of all profiles in each cluster are shown as colored density distributions: red (blue) for representative profiles that are asymmetric towards long (short) wavelengths, green for symmetric profiles. The asymmetry is calculated by considering the profile integration long-ward and short-ward from the profile peak. The total number of profiles is $N = 74480$.}
        \label{fig:kmeans}
    \end{figure*}

        Besides limiting the region where the training pixels were chosen from, we also made two prior criteria for including pixels in the training set.
        One criterion was on the strength of the line: if a profile had emission strength below 25~counts~s$^{-1}$, we did not include it in the training set unless it was above a certain threshold in full width at half maximum (FWHM). 
        We included a pixel that did not meet the first criterion in the training set if the FWHM was greater than 69~$\kms$. 
        After defining our training set, we used $k = 99$ clusters for training the model.
        After training our model, we appended the background profile, i.e., the average profile of all pixels in the dataset, excluding when IRIS data was heavily affected by cosmic ray hits during passage through the South Atlantic Anomaly, to the final $k$-means model with 100 clusters.
        This last profile we added to ensure that the quieter spectral profiles did not get picked up by the cluster centres involving the weaker profiles.
        Figure~\ref{fig:kmeans} shows all the cluster representative profiles ordered by peak intensity. 
        It presents the wide variety in complex profiles related to the UV burst.

\FloatBarrier

\section{IRIS SJI 1400 \AA\ UV burst light curve}
\label{appendix:sji1400_lightcurve}

\begin{figure}[h!]
    \centering
    \includegraphics[width=\columnwidth]{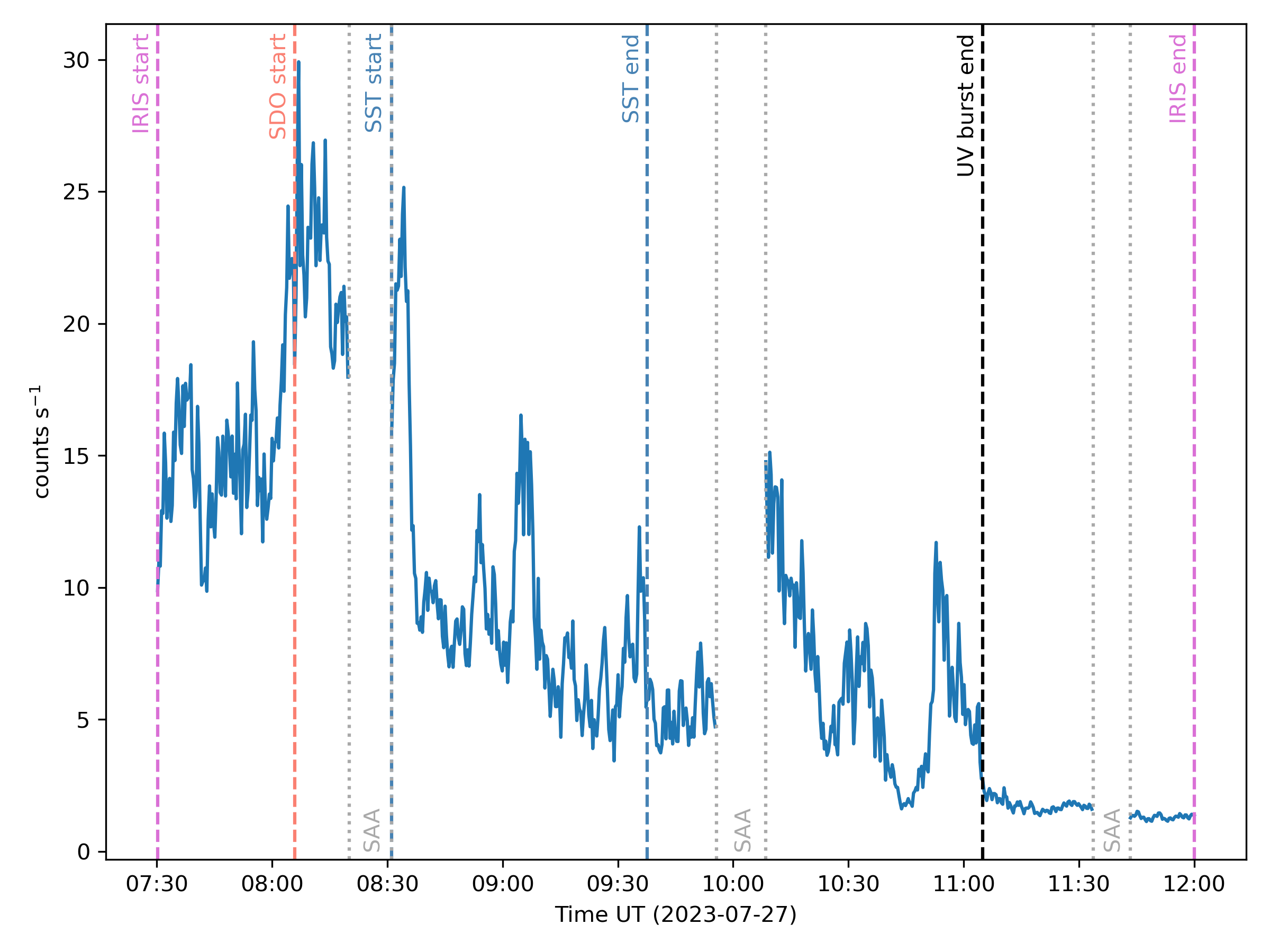}
    \caption{Light curve of the UV burst in IRIS SJI 1400~\AA, measured as the average signal over an area of 4\farcs 0$\times$4\farcs 0 centered on the UV burst. Measurements during the times IRIS passed through the South Atlantic Anomaly (SAA) are strongly affected by noise and are not included, these periods are marked with vertical dotted lines. Vertical dashed lines mark the start and end times of the IRIS and SST observations, the start of the SDO observations, and the end of the UV burst.}
   \label{fig:appendix_sji1400_lightcurve}
\end{figure}

    The light curve of the UV burst presented in Fig.~\ref{fig:appendix_sji1400_lightcurve} was constructed by averaging over a small square centered on the UV burst. 
    The light curve clearly shows that the UV burst was active when IRIS began observing at 07:30~UT and remained active until it went quiet at $\sim$11:05~UT, as marked by the dashed black line, yielding a minimum lifetime of 3~h and 35~min. 
    The repeated peaks are consistent with recurrent energy release from episodic magnetic flux emergence, as discussed in Sects.~\ref{res: uv_burst} and \ref{dis: flux_em}.

\section{SDO/AIA filter ratios' dependence on temperature}
\label{appendix:filter-ratio}

\begin{figure}[h!]
    \centering
    \includegraphics[width=\hsize]{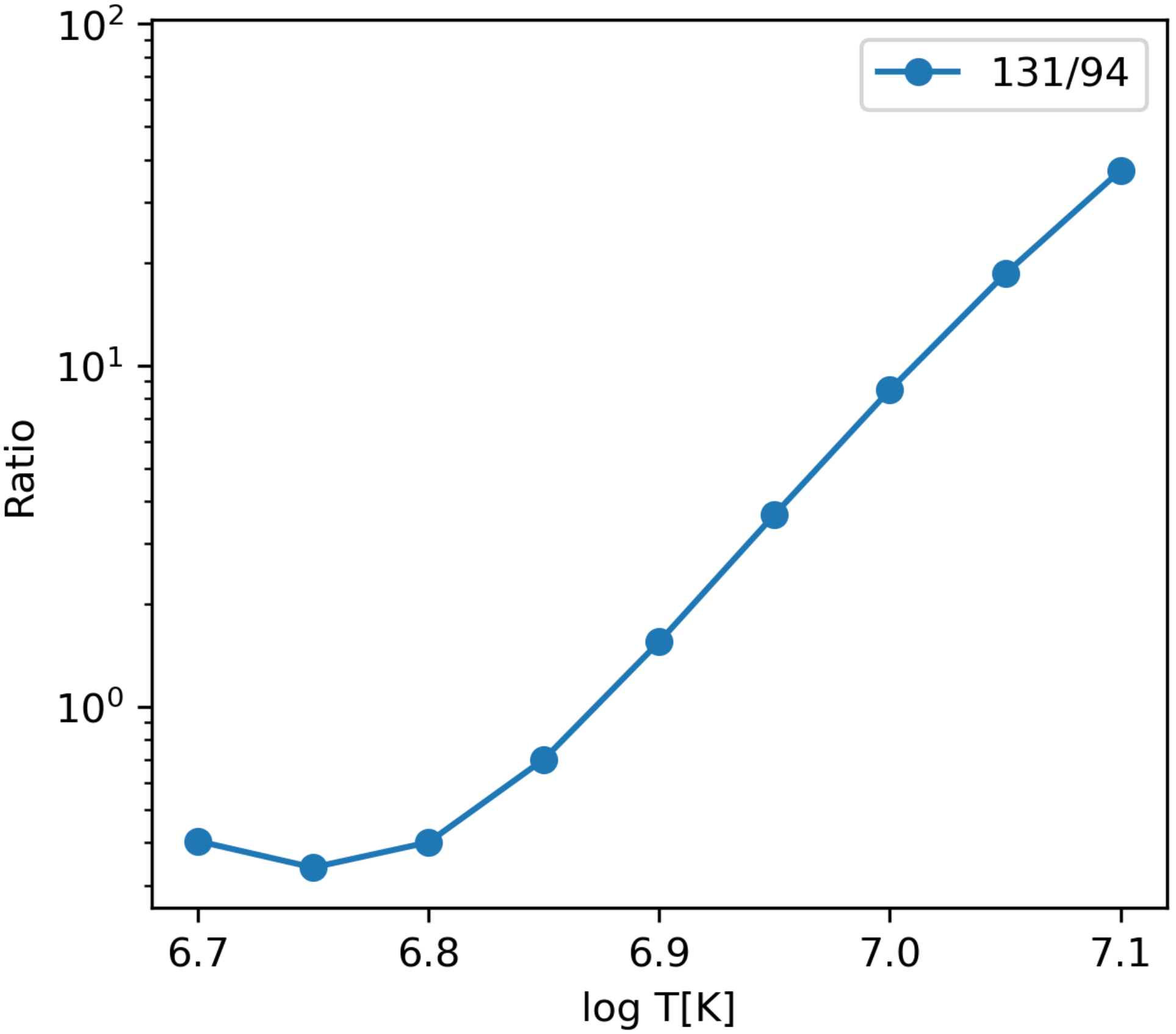}
    \includegraphics[width=\hsize]{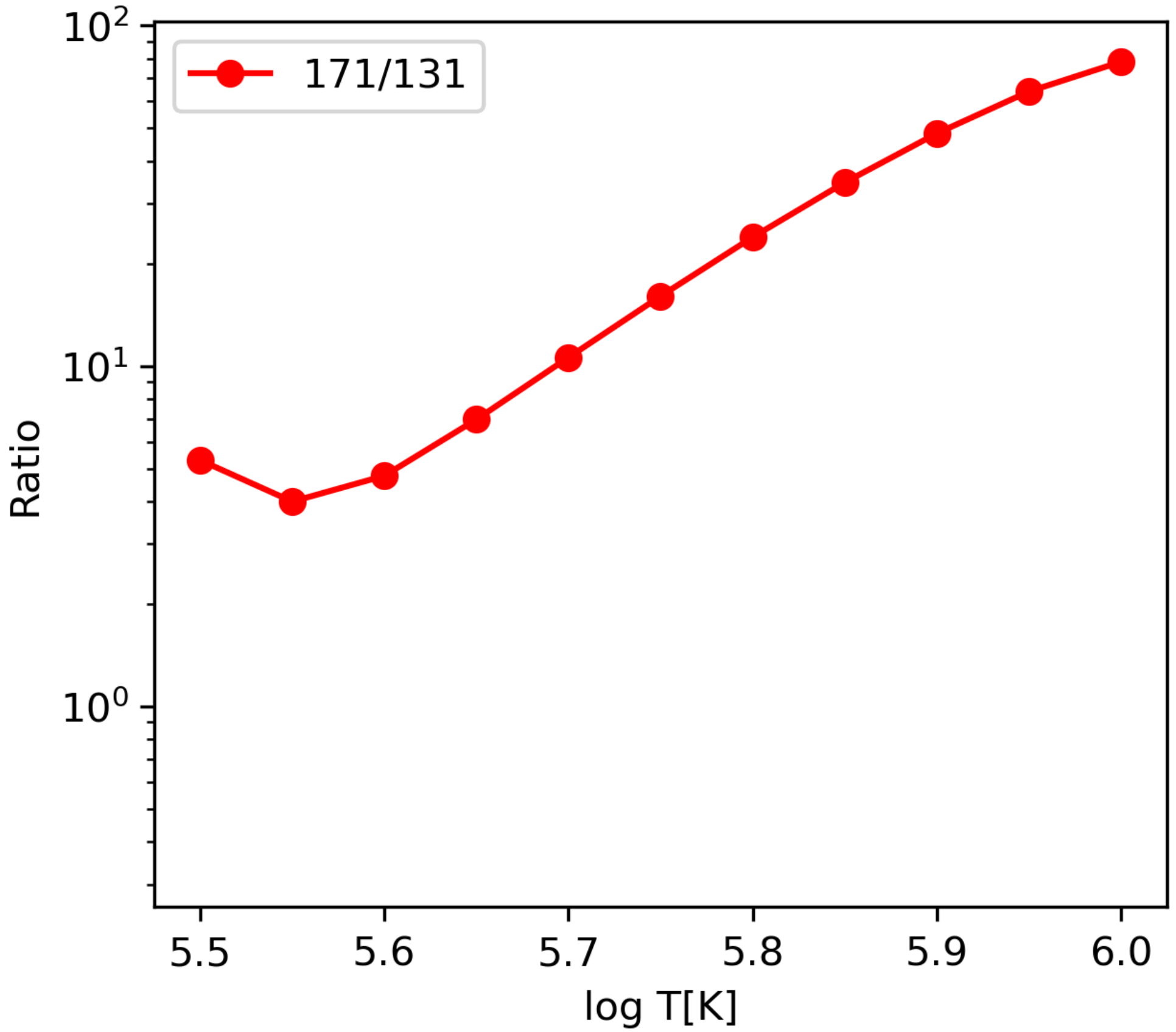}
    \caption{Ratios of the temperature response of the SDO/AIA 131 and 94 passbands between $\log T[\mathrm{K}] \in [6.75,7.1]$ (top panel), and 171 and 131 passbands between $\log T[\mathrm{K}] \in [5.5,6]$ (bottom panel). }
   \label{fig:appendix_filter_ratio}
\end{figure}

The filter-ratio technique relies on the fact that, under the optically thin and (nearly) isothermal assumption, the ratio of observed intensities in two AIA EUV channels can be mapped to an effective plasma temperature through the corresponding temperature response functions \citep{2011SoPh..269..169N}. Figure~\ref{fig:appendix_filter_ratio} shows the ratios of the AIA temperature responses, $R_{131/94}(T)=K_{131}(T)/K_{94}(T)$ and $R_{171/131}(T)=K_{171}(T)/K_{131}(T)$, over the temperature ranges relevant to our analysis. The $131/94$ curve exhibits a monotonic branch over $\log T[\mathrm{K}]\approx 6.8$--$7.1$, enabling a single-valued inversion from the measured $I_{131}/I_{94}$ ratio to an effective temperature in the hot regime, whereas $171/131$ provides sensitivity to ``cooler'' coronal plasma around $\log T[\mathrm{K}]\approx 5.6$--$6.0$. We include these curves to document the temperature dependence and to justify the choice of ratio-temperature mapping interval used in the main text.

 For the present work we take the response function from Solarsoft IDL, obtained with the command \verb|aia_get_response.pro| with the \verb|chiantifix| and \verb|eve| keywords. The `\verb|chiantifix|' keyword applies an empirical correction to the AIA~94, and AIA~131, channels to account for emission not included in CHIANTI \citep{2011ApJ...732...81A}\footnote{\url{https://soho.nascom.nasa.gov/solarsoft/sdo/aia/response/chiantifix_notes.txt}}, while the `\verb|eve|' keyword provides good agreement with the Extreme Ultraviolet Variability Experiment (EVE) of SDO. The code calculates the response functions using version 10.1 of the CHIANTI atomic database \citep{1997A&AS..125..149D,2012ApJ...744...99L, 2021ApJ...909...38D}.

\FloatBarrier

\section{XRT Constraints on the Coronal Contribution of the UV Burst}
\label{appendix:xrt_cut_off}

\begin{figure*}[h!]
    \centering
    \includegraphics[width=\linewidth]{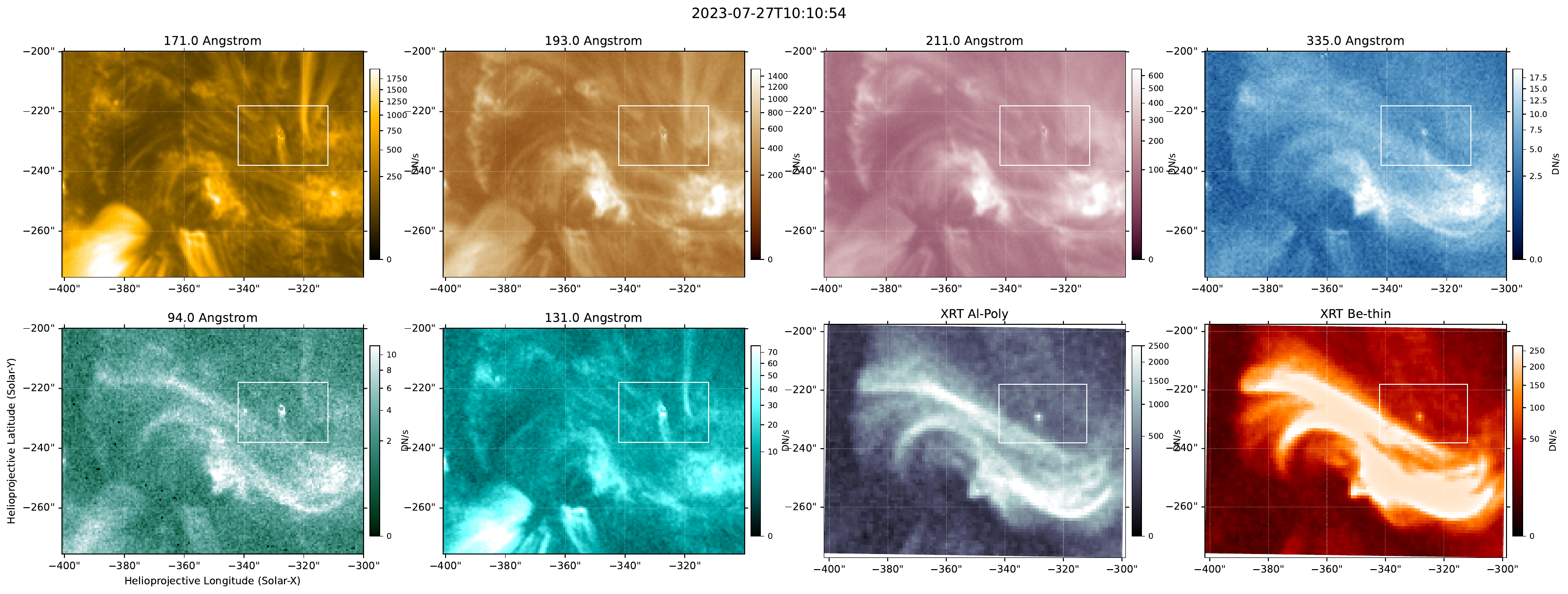}
    \caption{Co-aligned AIA and Hinode/XRT views of the active region at 2023-07-27 10:10:54~UT, ordered approximately by increasing sensitivity to hotter plasma from left to right and top to bottom. The panels show AIA 171, 193, 211, and 335~\AA\ in the top row, and AIA 94 and 131~\AA, followed by XRT Al\_poly and Be\_thin, in the bottom row. The white rectangle marks the common burst+dome field of view used for the aperture-based AIA--XRT comparison. The compact EUV burst is most clearly seen in AIA 94 and 131~\AA, while associated emission is also detected in the XRT filters, supporting the presence of plasma at coronal to multi-MK temperatures within the same localized structure.}
   \label{fig:xrt_aia_common}
\end{figure*}

\begin{figure*}[t]
    \centering

    \begin{subfigure}[t]{0.48\textwidth}
        \centering
        \includegraphics[width=\linewidth]{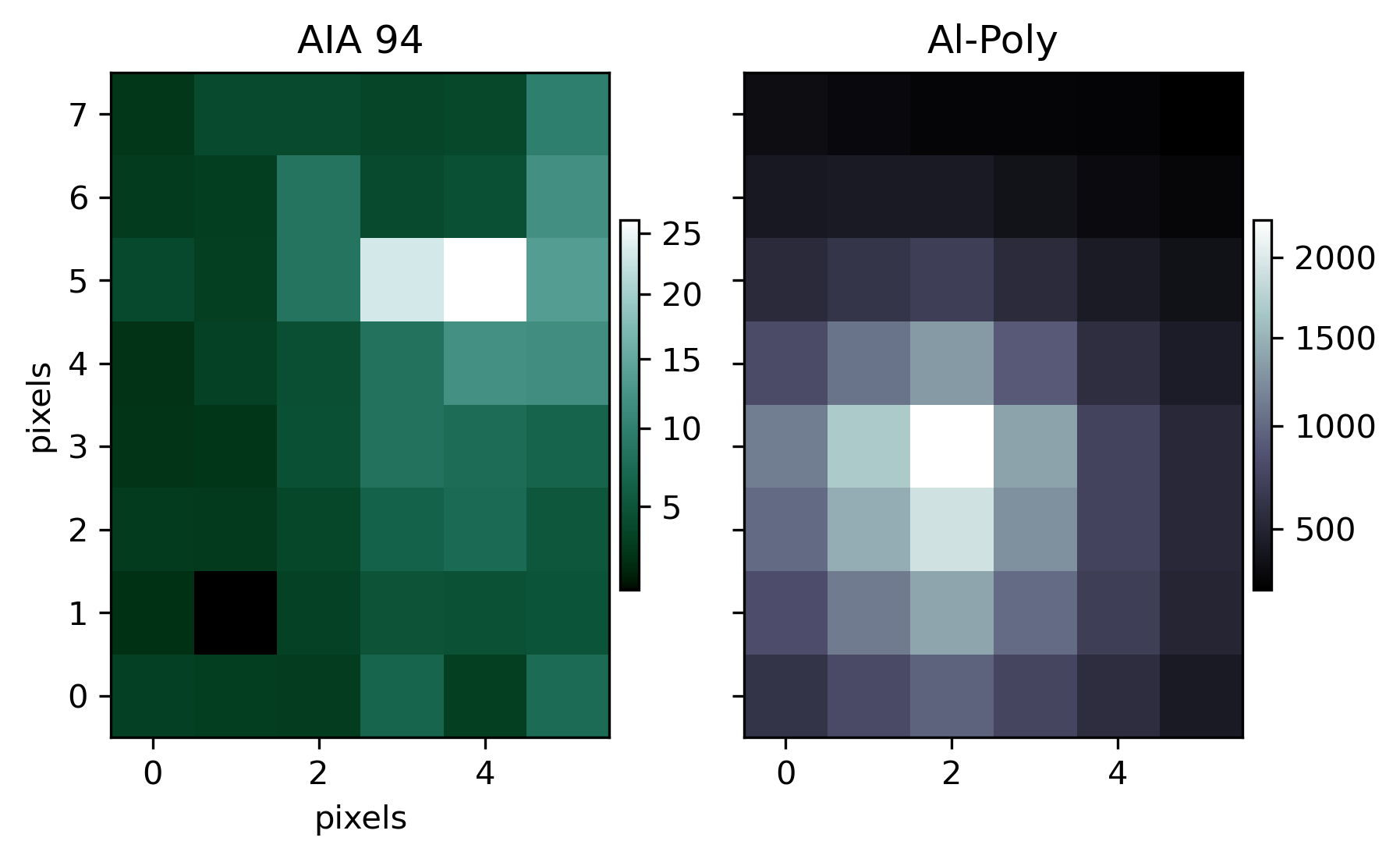}
        \caption{}
        \label{fig:common_omega}
    \end{subfigure}
    \hfill
    \begin{subfigure}[t]{0.48\textwidth}
        \centering
        \includegraphics[width=\linewidth]{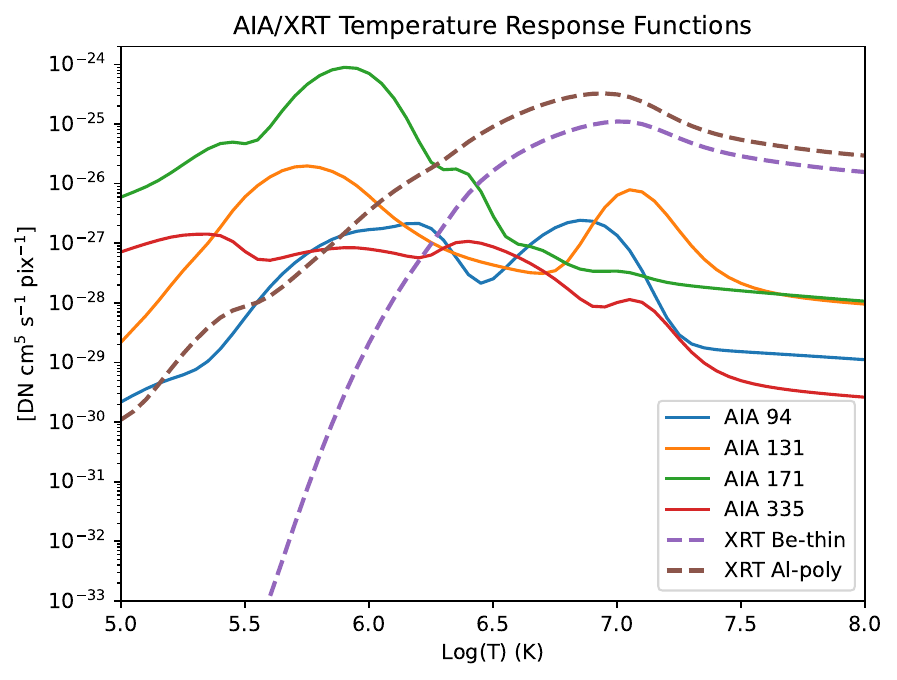}
        \caption{}
        \label{fig:aia_xrt_rfs}
    \end{subfigure}

    \caption{AIA/XRT setup used to constrain the lower-temperature cutoff of the burst-associated coronal emission. Panel~(a) shows the common burst+dome field of view used for the aperture-based comparison between the co-aligned AIA~94~\AA\ and XRT emission. Panel~(b) shows the temperature response functions of the relevant AIA and XRT filters, whose different sensitivities make the combined AIA/XRT comparison more constraining than the AIA-only DEM and filter-ratio diagnostics.}
    \label{fig:combined_common_rf}
\end{figure*}

To further assess whether the burst-associated signal seen in the AIA coronal channels represents genuine coronal plasma, we compared the event with coordinated Hinode/XRT observations. Figure~\ref{fig:xrt_aia_common} shows the larger common AIA/XRT context of the active region, while Fig.~\ref{fig:common_omega} isolates the common field of view used for the burst-centered comparison. The XRT signal is co-spatial with the compact burst environment, supporting the interpretation that at least part of the burst/dome system likely reaches coronal temperatures.

\subsection{Lower temperature cut-off from the AIA 94~\AA\ and XRT observations}
\label{appendix:lower_T_cutoff}

To estimate a conservative lower temperature threshold for the burst-associated dome, we compared the observed XRT signal with the XRT flux predicted from the co-spatial AIA 94~\AA\ emission under an isothermal assumption \citep[as in][]{2025ApJ...983L...7B}. The analysis was carried out on the common AIA--XRT field of view enclosing the burst core and the underlying dome, after co-alignment of the datasets onto the same helioprojective grid (Fig.~\ref{fig:xrt_aia_common}).

For each trial temperature $T$, we estimated the isothermal emission measure from the AIA 94~\AA\ intensity according to
\begin{equation}
EM(x,y;T) = \frac{I_{94}^{\mathrm{exc}}(x,y)}{R_{94}(T)},
\end{equation}
where $I_{94}^{\mathrm{exc}}(x,y)$ is the background-subtracted AIA 94~\AA\ count rate at pixel $(x,y)$ and $R_{94}(T)$ is the AIA 94~\AA\ temperature response function. In the absence of background subtraction, $I_{94}^{\mathrm{exc}}$ reduces to the observed count rate $I_{94}$, in which case the inferred temperature should be regarded as an upper-limit consistency check on the total line-of-sight emission rather than on the burst excess alone.

The corresponding XRT count rate expected from the same isothermal plasma is then
\begin{equation}
I_{\mathrm{XRT}}^{\mathrm{pred}}(x,y;T)
=
EM(x,y;T)\,R_{\mathrm{XRT}}(T)
=
\frac{I_{94}^{\mathrm{exc}}(x,y)}{R_{94}(T)}\,R_{\mathrm{XRT}}(T),
\end{equation}
where $R_{\mathrm{XRT}}(T)$ is the temperature response of the relevant XRT filter. Since the burst and dome are not strictly co-spatial at the pixel level in AIA and XRT, we did not rely on a pixel-by-pixel equality test. Instead, we integrated the predicted and observed signals over the common aperture $\Omega$ (as in Fig.~\ref{fig:common_omega}) enclosing the burst+dome structure:
\begin{equation}
F_{\mathrm{XRT}}^{\mathrm{pred}}(T)
=
\sum_{(x,y)\in\Omega}
\left[
\frac{I_{94}^{\mathrm{exc}}(x,y)}{R_{94}(T)}\,R_{\mathrm{XRT}}(T)
\right],
\end{equation}
and compared this quantity with the observed aperture-integrated XRT flux
\begin{equation}
F_{\mathrm{XRT}}^{\mathrm{obs}}
=
\sum_{(x,y)\in\Omega}
I_{\mathrm{XRT}}^{\mathrm{exc}}(x,y).
\end{equation}

\begin{figure}[!h]
    \centering
    \includegraphics[width=\hsize]{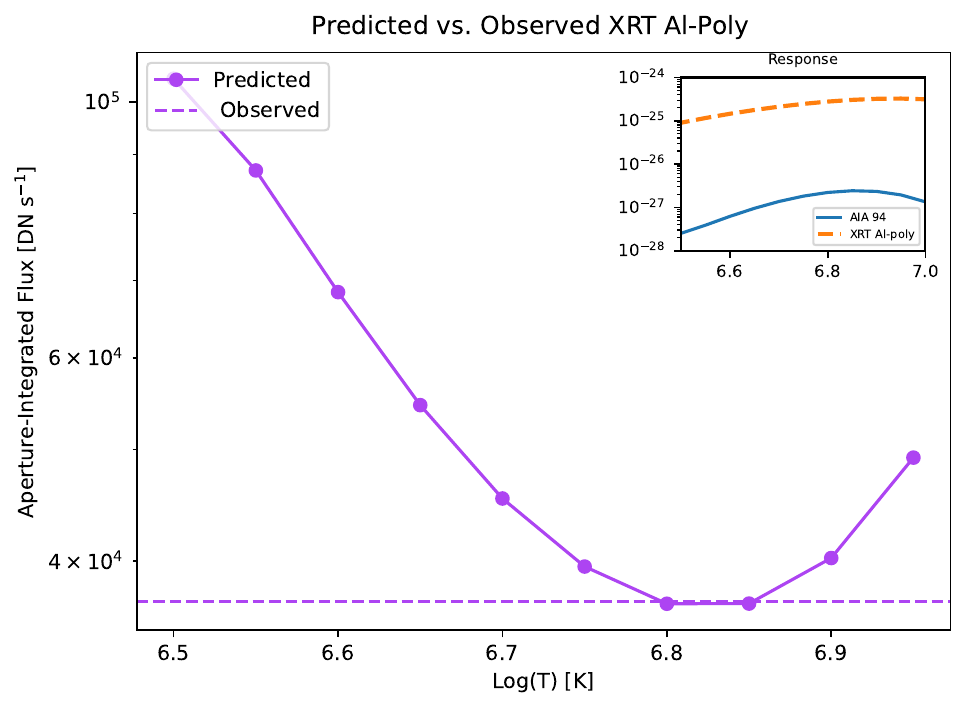}
    \caption{Aperture-integrated XRT Al\_poly flux predicted from the AIA 94~\AA\ emission under an isothermal assumption, shown as a function of temperature for the common burst+dome aperture. The solid curve gives the predicted XRT Al\_poly flux obtained by converting the AIA 94~\AA\ intensity into an isothermal emission measure and folding it through the XRT Al\_poly temperature response, while the dashed horizontal line indicates the observed aperture-integrated XRT Al\_poly flux. The lower intersection between the predicted and observed curves defines the conservative lower temperature cut-off, which in this case occurs near $\log T \approx 6.8$, indicating plasma in the multi-MK regime. The inset shows the AIA 94~\AA\ and XRT Al\_poly temperature response functions over the relevant temperature range.}
    \label{fig:pred_obs_xrt_flux}
\end{figure}

The variation of the predicted and the observed XRT~Al-poly aperture-integrated flux is shown in Fig.~\ref{fig:pred_obs_xrt_flux}. The lower temperature cut-off was defined as the lowest temperature at which the predicted XRT flux equals the observed one,
\begin{equation}
F_{\mathrm{XRT}}^{\mathrm{pred}}(T_{\min}) = F_{\mathrm{XRT}}^{\mathrm{obs}}.
\end{equation}
In practice, the predicted Al\_poly curve shows a shallow minimum, so that two intersections with the observed Al\_poly flux may occur. In that case, we adopt the lower-temperature intersection as a conservative lower-limit estimate. This procedure does not yield a unique isothermal temperature, but rather the minimum temperature for which the AIA 94~\AA\ emission can account for the observed XRT flux.

Applying this method to the common burst+dome aperture, we find that the AIA 94~\AA--XRT Al\_poly comparison is consistent with plasma in the multi-MK regime, with the lower intersection occurring near $\log T \approx 6.8$. The XRT Be\_thin filter, however, does not yield a simultaneous intersection, indicating that the structure is not described by a single self-consistent two-filter isothermal solution. We therefore interpret the value derived from Al\_poly as a lower-limit or consistency constraint, rather than as a unique temperature determination.

\newpage

\section{Surge evolution and burst visibility}
\label{appendix:surge}

\begin{figure*}[bh!]
    \centering
    \includegraphics[width=\linewidth]{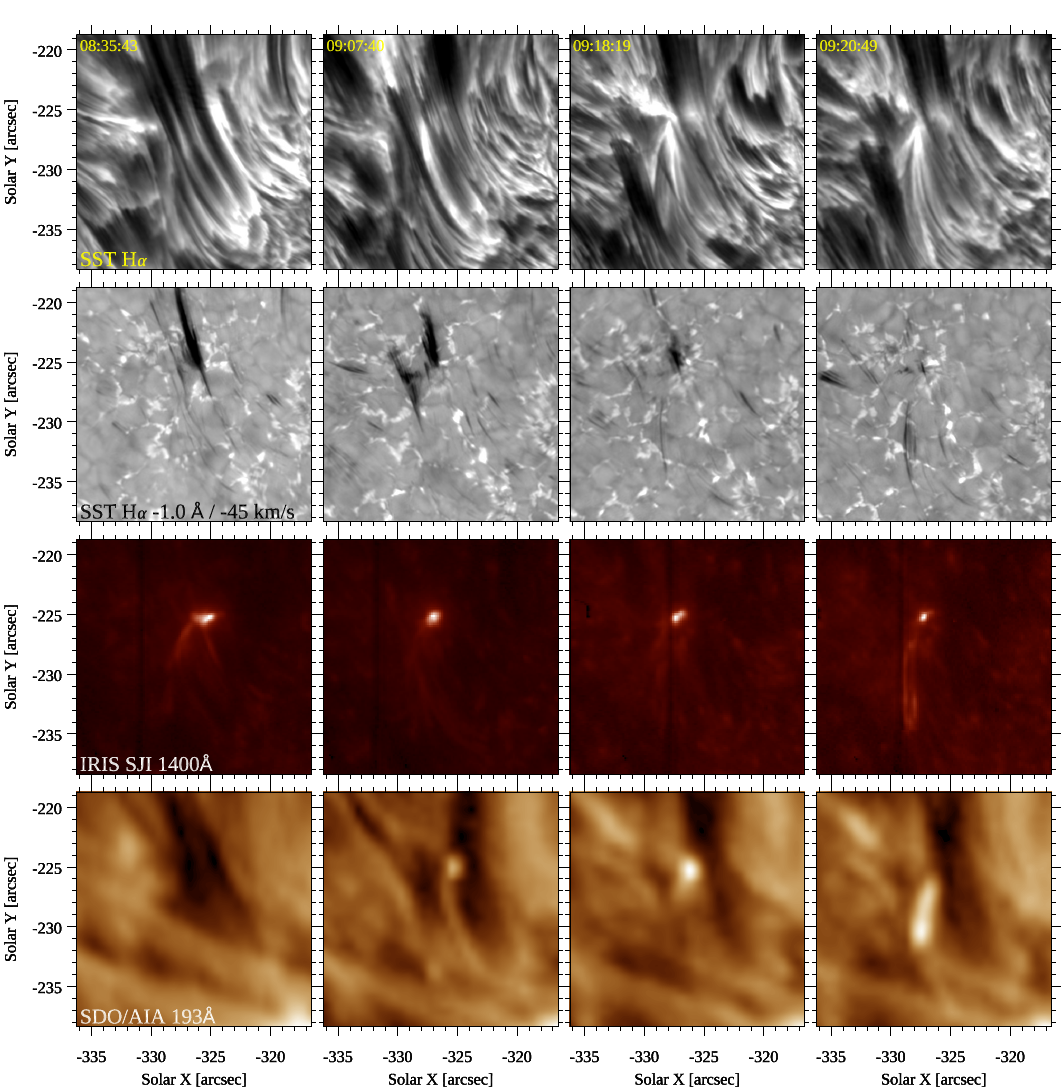}
    \caption{Surge evolution and visibility of the (E)UV burst. 
    The four columns show four different time steps in the time series, from top row to bottom: \Halpha\ line core, \Halpha\ wing, SJI 1400~\AA, and AIA 193~\AA. The time stamps are given in the top left of the top row. 
    The selected time steps are referred to in Sect.~\ref{dis: dome}.
    The FOV is the same as in Fig.~\ref{fig: structure}.
    The associated movie is available \href{http://tsih3.uio.no/lapalma/subl/mosand_uvb/mosand_uvb_figE1.mp4}{online}. 
    }
   \label{fig:surge}
\end{figure*}

Figure~\ref{fig:surge} shows different time steps in the observations illustrating varying visibility of the burst in AIA 193~\AA, while the burst is continuously visible in SJI~1400~\AA. 
It appears that the surge is covering the burst at times. Cold, dense plasma in the surge can absorb
the EUV emission due to the presence of neutral hydrogen and helium for wavelengths short ward of 912~\AA\ \citep{Anzer_Heinzel:2005}. 
The surge in \Halpha\ line core images is more extended than in \Halpha\ wing and covers a large part of the dome. 
In the \Halpha\ (and \Hbeta) wing, we only see parts of the surge that have sufficient velocity component along the line of sight. 
The surge has largest extension in \Halpha\ line core and in the AIA EUV images (also see Fig.~\ref{fig: overview}).  

\end{appendix}
\end{document}